\documentclass[journal, twocolumn, 10pt]{IEEEtran}
\usepackage{cite}
\usepackage{url}
\usepackage{amsmath,amssymb,amsfonts}
\usepackage{mathtools}
\DeclarePairedDelimiter{\ceil}{\lceil}{\rceil}
\usepackage{algorithmic}
\usepackage{graphicx}
\usepackage[caption=false,font=footnotesize]{subfig}
\usepackage{multirow}
\usepackage{textcomp}
\usepackage{pifont}
\usepackage[utf8]{inputenc}
\usepackage[T1]{fontenc}
\usepackage{tabularx}
\newcolumntype{L}[1]{>{\raggedright\arraybackslash}p{#1}}
\newcolumntype{C}[1]{>{\centering\arraybackslash}p{#1}}
\newcolumntype{R}[1]{>{\raggedleft\arraybackslash}p{#1}}

\begin{document}

\title{Real-time Interference Identification via \\Supervised Learning: Embedding Coexistence Awareness in IoT Devices}
\author{{Simone Grimaldi},
{Aamir Mahmood},~\IEEEmembership{Member, IEEE}, and {Mikael Gidlund},~\IEEEmembership{Senior Member, IEEE}
\thanks{S. Grimaldi, A. Mahmood and M. Gidlund are with the Department of Information Systems and Technology, Mid Sweden University, Sundsvall, 851 70 Sweden (e-mail: firstname.lastname@miun.se)}
\vspace{-15pt}
}



\maketitle

\begin{abstract}
Energy sampling-based interference detection and identification (IDI) methods collide with the limitations of commercial off-the-shelf (COTS) IoT hardware. Moreover, long sensing times, complexity and inability to track concurrent interference strongly inhibit their applicability in most IoT deployments.
Motivated by the increasing need for on-device IDI for wireless coexistence, we develop a lightweight and efficient method targeting interference identification already at the level of single interference bursts. 
Our method exploits real-time extraction of envelope and model-aided spectral features, specifically designed considering the physical properties of signals captured with COTS hardware.
We adopt manifold supervised-learning (SL) classifiers ensuring suitable performance and complexity trade-off for IoT platforms with different computational capabilities.
The proposed IDI method is capable of real-time identification of IEEE 802.11b/g/n, 802.15.4, 802.15.1 and Bluetooth Low Energy wireless standards, enabling isolation and extraction of standard-specific traffic statistics even in the case of heavy concurrent interference.
We perform an experimental study in real environments with heterogeneous interference scenarios, showing 90$\mskip3mu\%$--97$\mskip3mu\%$ burst identification accuracy. Meanwhile, the lightweight SL methods, running online on wireless sensor networks-COTS hardware, ensure sub-ms identification time and limited performance gap from machine-learning approaches.
\end{abstract}



\section{Introduction}
\label{SEC:Intro}

\IEEEPARstart{I}{ternet} of things (IoT) is empowering massive connectivity of objects, machines and devices for realizing smart-home, -building and -industrial applications. In this respect, short-range radio technologies such as WLAN (IEEE 802.11), Bluetooth/BLE (IEEE 802.15.1) and Zigbee (IEEE 802.15.4) etc. are in pivotal position to provide the needed local-area connectivity in unlicensed bands \cite{CellularIoT}. However, relying on these already widely adopted heterogeneous technologies for massive IoT comes with a caveat of cross-technology interference. The interference is usually detrimental for performance in co-located and concurrent operation \cite{coexistence_gidlund, SDR_Emulator_WB} in unlicensed bands, especially when coexistence---detection, identification and avoidance---mechanisms are ignored. 
As the domain of IoT services expands, interference characteristics strongly diversify in time, frequency and space domain.
Therefore, each device must have a built-in intelligence to detect, classify and characterize interference in distributed manner, which we study in detail in this paper, such that an interference-source specific mitigation strategy can be devised.


In unlicensed bands, a \textit{de facto} form of agility to interference is based on benign clear channel assessment (CCA). CCA can blindly---without knowing the source---detect interference and defer transmissions. However, CCA is unfavorable, in terms of medium access opportunities, especially to low-power systems as 802.15.1, BLE and 802.15.4~\cite{IoT_Era_Coexistence}. The other naive countermeasures are time-slotted channel hopping, manual channel blacklisting \cite{WHART} and link-quality estimation \cite{ISA100}, which are best effort and/or lazy to react to interference. As a result, many recommendations (e.g., IEC 62657-2 \cite{IEC}) suggested the adoption of interference-aware transmission (IAT) schemes in order to meet any quality of service (i.e., reliability and timeliness) requirements of diverse IoT applications.

A fundamental block for any IAT scheme is interference detection and identification (IDI). In the literature, a common approach to IDI is energy sampling (ES)-based interference detection followed by feature-based identification. This approach is usually straightforward and implementable in commercial off-the-shelf (COTS) hardware with radios of limited time/frequency sampling resolution. However, ES-based identification generally requires a sufficiently-long sampling time, mandating the root radio network (RRN) to defer its routine operation, while the storage size and processing of large set of samples leads to cumbersome and inefficient IDI implementation. Moreover, the detection of concurrent interference from multiple interfering radio networks (IRNs) is generally only possible with dedicate hardware, making it non-scalable for massive IoT deployments.

In this paper, mindful of these gaps, we present a real-time and lightweight solution to IDI in ISM bands (using 2.4$\mskip3mu$GHz as an example) that can differentiate among heterogeneous wireless technologies appearing in isolation or concurrently. By combining signal bandwidth and envelope information, and their slender extraction in COTS hardware (using 802.15.4 radio), with the intelligently tailored supervised-learning (SL) classification-trees, our solution enables on-board burst-based interference identification, predominantly in real-time. Our main contributions can be summarized as:
\begin{enumerate}
\item We develop a real-time burst-based interference identification solution for massive IoT environments, suitable for COTS hardware, which to authors' best knowledge is the first IDI method of this kind.

\item We bring the identification time (the time for detecting and processing an interference burst) to minimum, with respect to minimum interference-to-noise ratio (INR) and on-air-time (OAT) achievable with the employed COTS platform.

\item Apart from IDI, our solution provides a first such framework based on COTS hardware that allows onboard inference of the traffic distributions of concurrent heterogeneous IRNs, desired by coexistence solutions exploiting channel idle times~\cite{Petrova_IdleTime}.

\item The proposed method, instead of flimsy and heuristic power threshold-based features, utilizes signal features with unrestrictive requirement in reference to actual noise floor. While, we scrutinize the impact of INR on the identification performance.

\item We develop an analytical model for the key-enabler spectral features (SFs), which leads to an upper bound on classification gain and helps to fine-tune the SFs' parameters.

\item We compare the performance of SL classifiers of heterogeneous complexity and investigate the trade-off between implementable lightweight classifiers and complex machine-learning-based approaches.



\end{enumerate}
The rest of the paper is organized as follows. Section \ref{SEC:Background} gives the necessary background on IDI in the 2.4$\mskip3mu$GHz ISM-band, and discusses the related works. Section \ref{SEC:Proposed_method} describes the proposed method---including the feature extraction process and the classification strategies. Section \ref{SEC:Experimental_setup} presents the experimental setup, while Section \ref{SEC:Results} evaluates the performance of our IDI method. Section \ref{SEC:Discussion} investigates the implications of SFs and develops a analytical model for estimating the upper bound on the classification gain. Section~\ref{SEC:Discussion1} presents a use case of the proposed method. Finally, we conclude this work in Section \ref{SEC:Conclusion}.

\section{Background} \label{SEC:Background}
In this section, we develop the necessary background on heterogeneous characteristics of wireless technologies operating in ISM-bands, and the limitations of spectrum sensing in COTS hardware and related IDI methods in the literature.

\subsection{Coexistence of Wireless Technologies at 2.4$\mskip3mu$GHz}
In 2.4$\mskip3mu$GHz, wireless coexistence---often harmful---results from ubiquity of networks and devices employing IEEE 802.11 (with variants as b/g/n/ac), IEEE 802.15.1 (Bluetooth classic and low-energy BLE extension), and IEEE 802.15.4 standards. By comparing the PHY and medium access parameters in these standards (see Table \ref{TAB:Technologies_coex}), it is instantly noticeable that the utilized channel bandwidth and transmit power in 802.11 radios can be inundating for low-power standards. The same is validated by several studies (see \cite{coexistence_gidlund} and the references therein), which show that a co-located 802.11 pose serious concerns for reliability in 802.15.4-based WSNs, while the interference from 802.15.1 is less pronounced due to its  sub-{ms} fast-frequency-hopping (FFH) scheme and a channel blacklisting (CB) policy.
\bgroup
\begin{table}[h!]
\caption{Salient Features of Wireless Technologies in the 2.4$\mskip3mu$GHz ISM-band--A Perspective on Coexistence.}
\centering
\setlength\tabcolsep{3pt} 
\begin{tabular}{L{1.77cm}C{1.4cm}C{1.4cm}C{1.4cm}C{1.5cm}}
\hline
\textbf{Feature}&\textbf{802.15.1}&\textbf{BLE}&\textbf{802.15.4}&\textbf{802.11b/g/n}\\
\hline
{Channels (CH)}& 79 & 40 & 16  & 14 \\
{CH Numbering}& $[0,78]$ & $[0,39]$ & $[11,26]$   & $[1,14]$  \\
{CH Width}&1$\mskip3mu$MHz & 2$\mskip3mu$MHz & 5$\mskip3mu$MHz  & 20/40$\mskip3mu$MHz\\
{Data rate}& 1,2,3$\mskip3mu$Mbps & 1$\mskip3mu$Mbps & 0.25$\mskip3mu$Mbps & $<600\mskip3mu$Mbps \\
{Modulation}& GFSK,*-DPSK & GFSK & OQPSK & Several \\
{Tx Power}& $\leq 20 \mskip3mu$dBm & $\leq 10 \mskip3mu$dBm& $\leq 0 \mskip3mu$dBm & $\leq 20 \mskip3mu$dBm\\
{Coexistence}& FFH, CB & FFH, CB & ED-CCA & ED-CCA\\
{Diffusion}& \ding{108}\ding{108}\ding{108}& \ding{108}\ding{108}\ding{109}& \ding{108}\ding{109}\ding{109} & \ding{108}\ding{108}\ding{108}\\
\hline
\label{TAB: accuracy_global}
\end{tabular}
\label{TAB:Technologies_coex}
\vspace{-17pt}
\end{table}
\egroup

\subsection{Energy Sampling with 802.15.4 Hardware}
Energy sampling (ES) is a generic spectrum-sensing process for capturing information in a certain time and frequency RF-resource through sampling the current induced by electromagnetic radiation on the desired radio interface. To this end, the energy detector employed in COTS WSN transceivers is a low-cost solution as compared to dedicated spectrum analyzers or software defined radios (SDR). However, as the radio front-ends in COTS hardware are designed mainly for 802.15.4 standard-specific operations (e.g., CCA), the prerogatives of ES are partly met. That is, specification-compliant frequency response of the radio leads to sub-Nyquist sampling and limited frequency resolution. In addition, the energy measurements are available only in the form a received signal strength indicator (RSSI), i.e., a 8$\mskip3mu$bit, $T_r=128 \mskip3mu\mu$s-moving-average filtered version of the baseband power envelope \cite{802154}.

\begin{figure}[h]
\centering
\includegraphics[width=0.48\textwidth,clip, trim=0cm 0cm 0cm 0cm]{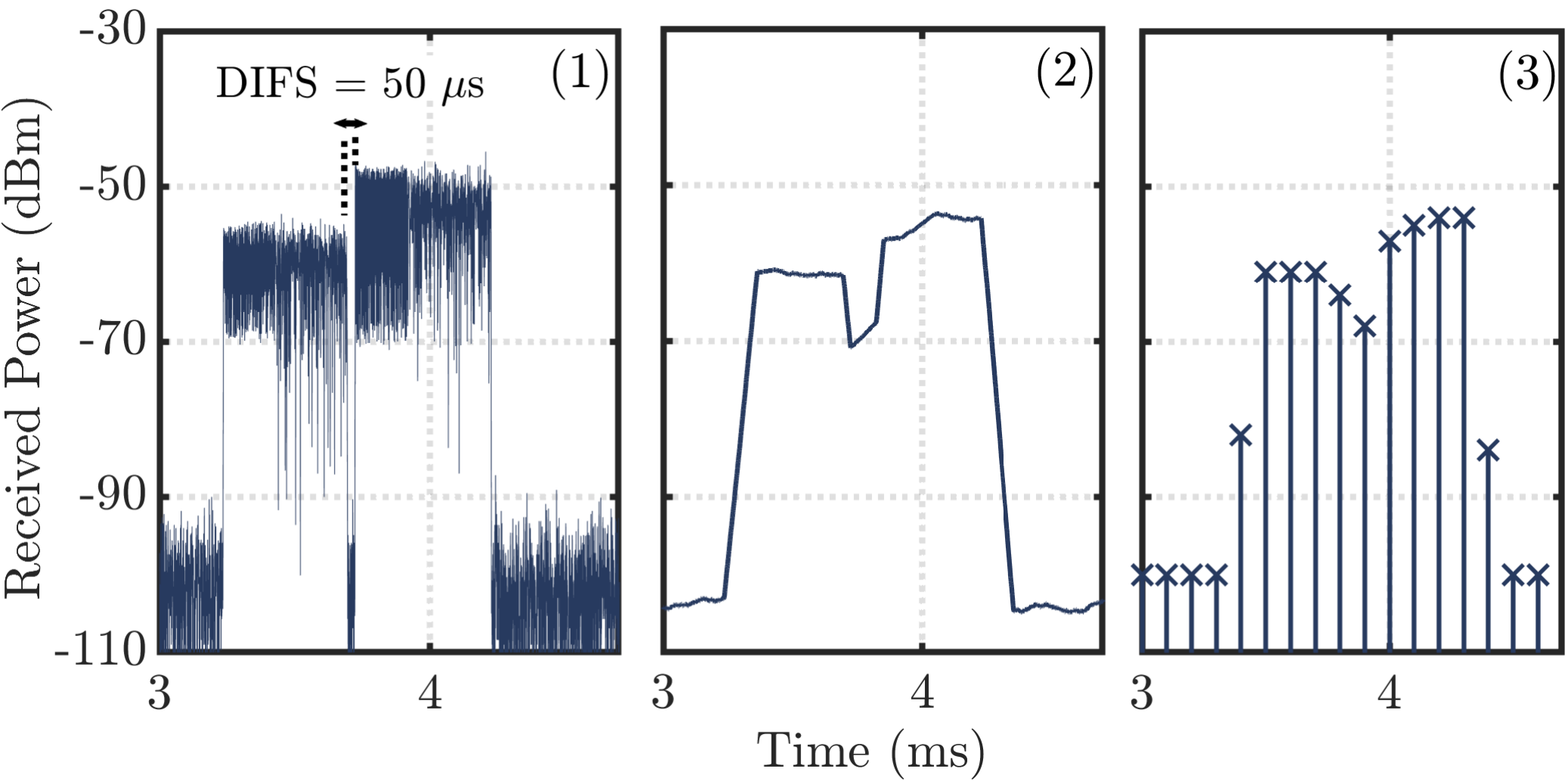}
\caption{Information loss in RSSI filtering: (1) power envelope of 802.11 bursts extracted from high-resolution I/Q data, (2) after 802.15.4-compliant moving average filtering, (3) sampling and quantization 8$\mskip3mu$bit/10$\mskip3mu$kHz---a common situation with COTS WSN hardware.}
\label{FIG:sensing_USRP_WSN}
\vspace{-10pt}
\end{figure}

In Fig.~\ref{FIG:sensing_USRP_WSN}, we show an example of information loss in RSSI calculation using a high-resolution I/Q trace of two consecutive 802.11g packets. The effect of digital RSSI filtering, here obtained by an 802.15.4-emulator, is most devastating in terms of information loss.  Due to low-pass filtering (LPF) effect, not only it wipes out the information on signal envelope---inhibiting modulation-based identification (such as \cite{deep_learning}), but also caps time-resolution---dampening the information gain of sampling frequencies above $f_r=7.8\mskip3mu$kHz. This, in turn, reflects on the inability to capture short interference bursts and inter-frame spaces, such as the 802.11 DCF inter-frame space (DIFS). Despite these limitations, it remains attractive to perform IDI using low-cost 802.15.4 hardware for enabling in-device distributed sensing and adaptation capability.

\subsection{Related Works}
\label{SUBSEC:Related_works}
ES-based IDI methods commonly rely on signal features extracted from RSSI samples. These features are subsequently mapped to an interference class by an interference classification engine, while the two most common classification techniques are technology-specific (heuristic) thresholding \cite{ChowdhurySpectral, Zach_hindawi, King_DCCA} and machine learning \cite{Grimaldi, multiSourceInterference}. 
In WSNs community, the interest has been towards low-complexity and low-cost solutions such that the interference identification is demodulation free.
This requirement, induced mainly by the hardware-constraints of COTS-radios, complicates the identification time and accuracy.
That is, usually long traces of low-resolution RSSI-based channel energy samples are required since the technology-specific temporal (e.g., idle/busy time and distributions, burstiness, periodicity etc.) and spectral (power distribution with respect to frequency channels) features become apparent only at large observation windows. For instance, temporal features-based identification in \cite{Zach_hindawi} requires observation time in seconds to achieve moderate accuracy.  In essence, this is not only due to hardware limitations but also to the selection of signal features which are limited in scope.

 



When it comes to the identification of multiple heterogeneous interference sources appearing concurrently, there is a limited work in the literature. 

In \cite{multiSourceInterference}, the identification of concurrent multi-source interference is based on $k$-means clustering of RSSI-samples. Using RSSI sampling of $21\mskip3mu$kHz and sampling window of $3\mskip3mu$s, the authors \cite{multiSourceInterference} in achieved a classification accuracy of 90$\mskip3mu\%$, which however reduces further if the 802.15.4 network is not silent during observation time. Although, IDI enhancement in the presence of 802.15.4 traffic is addressed in \cite{King_DCCA} using power variations in CCA, however, the overall detection performance reduces significantly.  


A different approach to IDI is to search for interference-specific bit error patterns in the received packets. In \cite{HermanSonic,Barac_diganostic}, for example, such patterns are exploited by mean of supervised-learning (SL) or algorithmic approach. While the identification time is rather limited (i.e., in the order of tens of ms with COTS hardware), the methods are constrained by the event of receiving an interfered RRN packet, generally leading to higher detection time as compared to ES methods. 

Using specialized hardware, such as WLAN cards and SDRs, for protocol-free IDI has also been investigated in many studies, e.g., \cite{RayanchuAirShark,Li_demodFree,deep_learning} and references therein. As this hardware can ensure high sampling frequency and high-resolution I/Q data, the benefit of more complex classifiers (e.g., deep learning-based~\cite{deep_learning}) increases and higher accuracy is generally achieved. However, we have shown that, even with limited sensing resolution and lightweight supervised learning, COTS IoT nodes can reach the same-level of accuracy in real-time.

An objective comparison of the related works with the proposed solution in this paper can be made from Table~\ref{tab:SummaryOfIDIInRelatedWorks}. It shows how reactive and accurate our solution is, while enabling identification of both concurrent heterogeneous IRNs (CI) and RRNs. 

\bgroup
\def\arraystretch{1.2}
\begin{table*}[t]
	\centering
		\caption{Summary of Interference Detection and Identification Techniques in Related Works.}
		\label{tab:SummaryOfIDIInRelatedWorks}
	\begin{tabular}{c l l l c c c c l}
    \hline
				\textbf{Ref.}  & \textbf{Data source} &\textbf{Features} & \textbf{Classification} & \textbf{CI} & \textbf{RRN} & \textbf{IDI time} & \textbf{Sampling rate} & \textbf{Accuracy}\\ \hline
                				Current & RSSI+CCA       & Envelope + Spectral & SL (CTs,SVM)      & \ding{51} & \ding{51} & 620$\mskip3mu\mu$s     & 18$\mskip3mu$kHz & 90$\mskip3mu\%$--97$\mskip3mu\%$ \\	


					\cite{Zach_hindawi} & RSSI       & Temporal pattern            & Algorithmic      & \ding{55} & \ding{55} & 700$\mskip3mu$ms   & 8$\mskip3mu$kHz & -- \\	
				
				\cite{multiSourceInterference} & RSSI       & Temporal pattern            &  UL   & \ding{51} & Limited & 3$\mskip3mu$s   & 21$\mskip3mu$kHz & 90$\mskip3mu\%$ \\	
				
				\cite{King_DCCA} & RSSI       & Temporal + Spectral            & Algorithmic      & \ding{55} & \ding{51} & 256$\mskip3mu\mu$s   & 31$\mskip3mu$kHz & 80$\mskip3mu\%$ \\	
				
					\cite{ansari_wispot} & RSSI (dual-radio)       & Spectral pattern              & Algorithmic    & Limited & \ding{55} & 310$\mskip3mu$ms   & 14$\mskip3mu$kHz & 96$\mskip3mu\%$ \\	
				
				\cite{HermanSonic}    & Received packets &  Bit-error, RSSI, LQI & SL (SVM) & \ding{55} &   \ding{55}     & 28$\mskip3mu$ms & -- & 73$\mskip3mu\%$--76$\mskip3mu\%$\\ 	
				
				\cite{Barac_diganostic}             & Received packets & Bit-error & Algorithmic & \ding{55} &   \ding{55}     & 30$\mskip3mu$ms & -- & 91$\mskip3mu\%$\\
				
							\cite{RayanchuAirShark}            & I/Q (Wi-Fi NIC) & Temporal + Spectral &  SL (CT) & \ding{51} &   \ding{51}     & 100$\mskip3mu$ms & 1$\mskip3mu$kHz--10$\mskip3mu$kHz & 91$\mskip3mu\%$--96$\mskip3mu\%$\\	
				
							\cite{Li_demodFree}            & I/Q (SDR) & Temporal + Spectral & Thresholding & \ding{51} &   \ding{51}     & 120$\mskip3mu$ms & $\sim \mskip3mu$MHz & 90$\mskip3mu\%$ \\  \hline
			\end{tabular}
\end{table*}
\egroup
\section{Proposed Method}
\label{SEC:Proposed_method}
In this work, contrary to earlier studies, we aim to stretch a hardware-limited WSN to perform burst-based IDI, enabling real-time identification of concurrent sources of interference. We additionally pursue the identification of packets transmitted by the 802.15.4 RRN, removing the need of idle period for spectrum sensing, for ensuring a IDI process with no performance impact on the RRN. 
To achieve these objectives, we optimize both the design of features and the classification strategy. In particular, we compensate the problem of time-resolution loss due to RSSI filtering process, via domain-switch. That is, we capture spectral features (SFs) within a burst duration to extract information on the bandwidth of the single bursts. On the classification side, we evaluate a number of SL methods to find a reasonable trade-off between complexity and classification accuracy.


\begin{figure}
\centering
\includegraphics[width=1\linewidth,clip, trim=0cm 0cm 0cm 0cm]{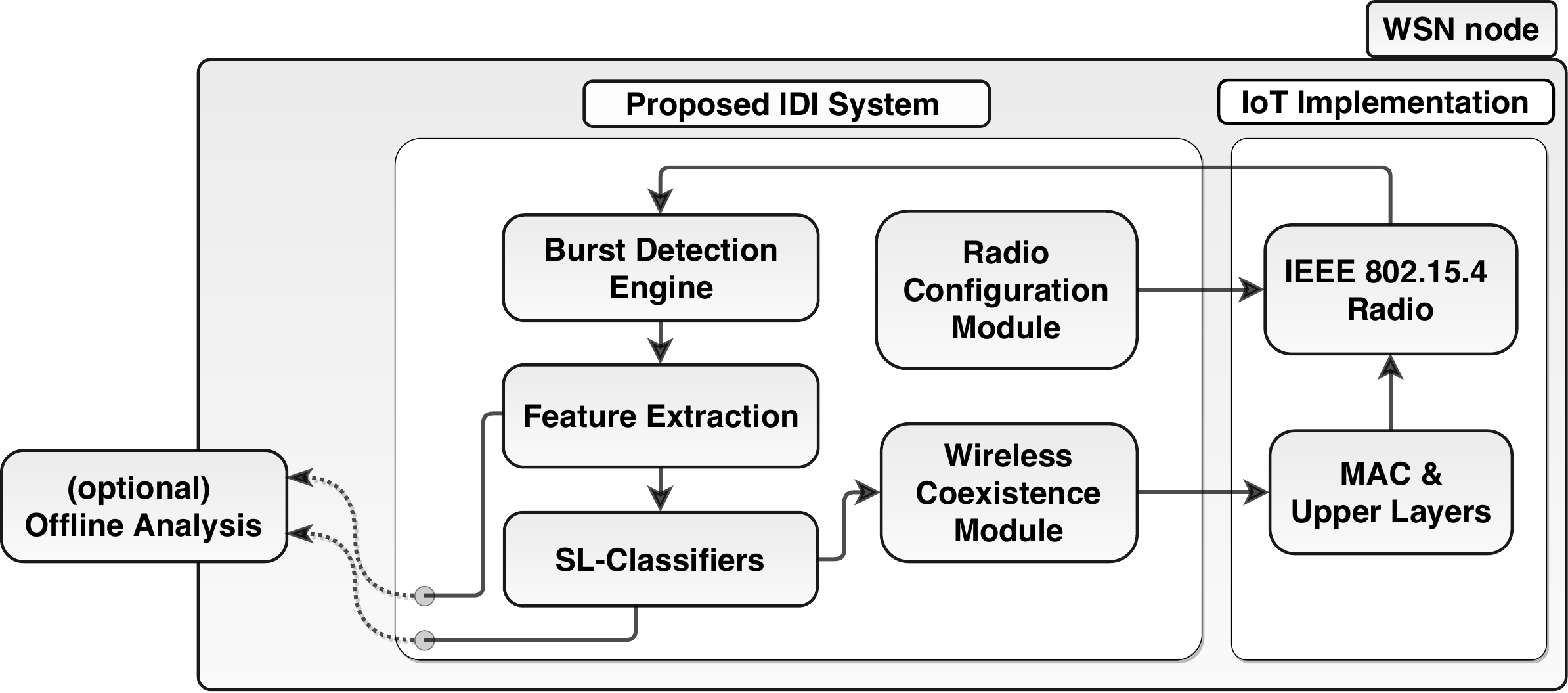}
\caption{Global scheme of the proposed method.}
\label{FIG:global_scheme}
\vspace{-12pt}
\end{figure}

\subsection{Interference Burst Detection}
The on-board burst detection engine samples the RSSI register with frequency $f_s = 18.5\mskip3mu$kHz, fetching 1$\mskip3mu$dB-resolution data, ensuring adequate super-Nyquist rate with respect to the RSSI LPF cutoff frequency $f_r$. Signal bursts are separated from noise in real-time using a threshold-based criteria with threshold $P_T = \mu_N+2 \sigma_N$, where $\mu_N$ and $\sigma_N$ mean and standard deviation of the AWGN noise due to the radio front-end, such that the probability noise-triggered bursts is minimized. Note that $\mu_N$ and $\sigma_N$ are device specific and usually provided by chip manufacturer, and can also be determined via a quick calibration process. Nevertheless, a conservative choice on $P_T$ only leads to a slight loss in detection sensitivity for low-INR bursts.

\subsection{Feature Extraction}
Upon burst detection, eight time- and frequency-domain features are extracted in real-time, which we describe below.

\subsubsection{Spectral Features}
\begin{figure}
\centering
\includegraphics[width=0.5\textwidth,clip, trim=1cm 3cm 0cm 1cm]{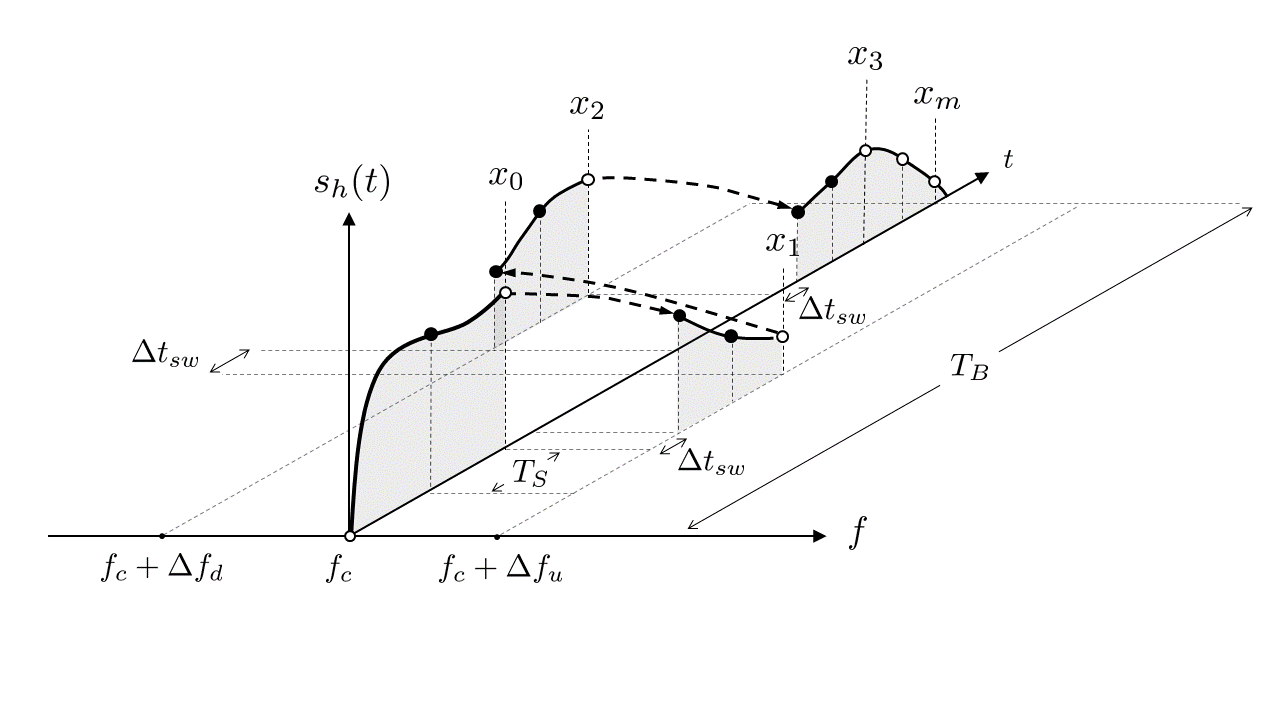}
\caption{Time-frequency representation of the employed intra-burst side-channel sampling method.}
\label{FIG:channel_swhitch_diagram}
\vspace{-12pt}
\end{figure}

The combined LPF effect of 802.15.4 channel- and RSSI calculation reduces the bandwidth of captured bursts to a few kHz, meaning that the spectral signature of different signals is completely removed. To overcome this limitation, the proposed method hunts for information in the frequency domain by using a simple, yet proactive strategy. Two fast upward and downward intra-burst frequency shifts are performed, and the RSSI values on the related side-bands are sampled and compared to the mean value of the burst envelope captured at the central frequency $f_c$. The extraction process of SFs is shown in Fig.~\ref{FIG:channel_swhitch_diagram}, depicting the collection of RSSI samples $x_n$ from the sampling of $f_r$-LPF-filtered version of the signal available at the radio interface.  The first sample $x_0$, representing the initial reading on the central band $f_c$, is acquired $\ceil*{{f_s}/{f_r}}=2$ samples after the first over-threshold reading. It ensures that the RSSI value has sufficient time to accommodate to the crest value of the burst. Subsequently, in order to perform a reading on the lower side-band, the on-board voltage controlled oscillator (VCO) is tuned to $f_c-\Delta f_d$. To reduce the correlation between the samples on the central and lower band, introduced by the moving average filter, the first two samples on the side-band are discarded so that $x_1$ reflects the value of the third sample. The same procedure is repeated to perform a reading $x_2$ on the upper side band $f_c+\Delta f_u$, and then the VCO is tuned back to $f_c$ to sample the remaining portion of the burst, until the last over-threshold sample $x_m$ is recorded.

Let $\mathbf{y} = \left[x_0, x_3,\cdots,x_m\right]$ be the set of samples collected at $f_c$ with mean $\bar{y}$ and cardinality $C_y=|y|$. Then the SFs with respect to upper side-band $F_{S_u}$ and lower side-band $F_{S_d}$ are simply defined as
\begin{equation}
F_{S_u}=\bar{y}-x_1;\: F_{S_d}=\bar{y}-x_2; \: F_{S_c}=c_Z
\label{EQ:SF_definition}
\end{equation}
where $F_{S_c}$ is a support feature, which reflects the channel number $c_Z\in[11,26]$ of 802.15.4 in reference to $f_c$.
In essence, the SFs exploit the native spectral differences among the coexisting signal families, under the condition that the frequency shift parameters $\Delta{f_u}$ and $\Delta{f_d}$ can guarantee sufficient separation among interference classes. The analysis leading to appropriate selection of these parameters is non-trivial, as we study in detail in Sec.~\ref{SSEC:Abalysis_of_SF} using a model-based approach, and leads to a symmetric selection of $\Delta f_u =\Delta f_d=2\mskip3mu$MHz, as shown in Table~\ref{TAB:sampling_parameters}.
\begin{table}
\caption{IDI-mode: Detail of Parameters for Burst Analysis.}
\centering
\resizebox{0.5\textwidth}{!}{\begin{tabular}{ccl}
\hline
{\textbf{Param.}}& {\textbf{Value}}&  {\textbf{Description}}\\
\hline
$D_r$ & -100$\mskip3mu$dBm--0$\mskip3mu$dBm/1$\mskip3mu$dB& RSSI: dynamic range/resolution\\
$f_c$ &  2405$\mskip3mu$MHz--2480$\mskip3mu$MHz &Central frequency (range) \\
$\delta_f$ &  1$\mskip3mu$MHz &Minimum frequency step\\
$\Delta f_{u,d}$ & $\pm$2$\mskip3mu$MHz & Side-channel offsets\\
$T_b$ & 324$\mskip3mu\mu$s--5000$\mskip3mu\mu$s&OAT of identifiable signal bursts\\
$T_s$ & 54$\mskip3mu\mu$s & Sampling period\\
$T_r$ & 128$\mskip3mu\mu$s & Period of moving average filter\\
$\Delta T_{sw}$ & 25$\mskip3mu\mu$s & Channel switching time\\
\hline
\label{TAB:sampling_parameters}
\end{tabular}}
\vspace{-15pt}
\end{table}

\subsubsection{Time- and Power-Domain Features}
We base the design of envelope features ($F_E$) on the inherent limitations of the COTS hardware observing that, while time- and power-domain information is scarce, macroscopic differences among packets from interference families remain observable. Specifically, since the information on the modulation format of the different signals and the effects of fast-fading is removed by the RSSI LPF, the observable variations of the 1$\mskip3mu$dB-quantized envelope are mainly due to:
a) the inability of the observation system to resolve two closely-spaced packets with interarrival time $T_I < T_r$, leading to the artifact of a single burst with steep envelope variation (see Fig.~\ref{FIG:sensing_USRP_WSN}), b) the short-term fading dynamics reflecting in slow RSSI variations in the order of few dB~\cite{Channel_gains} within the lifespan of the observed signal bursts (see Table~\ref{TAB:sampling_parameters}).

Under these premises, we introduce the following lightweight time- and envelope-features ($F_T$ and $F_E$):
\begin{itemize}
\item \textit{Burst length}: total sample length of the detected burst, hence, $F_{T_l}=C_y+6$.  
\item \textit{Burst mean power}: reflecting the mean envelope power extracted on the central frequency $f_c$, then, $F_{E_p}=\sum y/C_y$. 
\item \textit{Crest factor}: indicating the maximum envelope variation, i.e., dynamic range of the signal envelope, as, $F_{E_c}=\max(y)-\min(y)$,
\item \textit{Envelope ripple}: representing a measure of the maximum power variation between two consecutive samples.
\begin{equation}
F_{E_r}=\sum_{i=0}^{C_y-1} r\big(\vert y_{i+1}-y_{i}\vert\big) \; \text{ s.t.: } r(y)=
\begin{cases}
   1 \; \text{ if } \; y \geq P_E\\
   0 \; \text{ otherwise} \\
\end{cases}
\end{equation}
\end{itemize}
where $P_E$ is a threshold value, empirically set to 4$\mskip3mu$dB.
\subsubsection{CCA based feature}
In order to decrease the simultaneous channel access and packet collisions, the 802.15.4 standard mandates using one of the four CCA methods before making a medium access attempt \cite{802154}. The mode of particular interest is CCA Mode 2, which detects 802.15.4-compliant signal using the on-board OQPSK modem. We exploit CCA Mode 2 to acquire an identification marker of 802.15.4 signals. In this respect, a CCA Mode 2 is performed immediately after acquiring the first valid sample $x_0$, while the related binary feature $F_{\textrm{CCA}}\in [0,1]$ reflects the CCA outcome.

\subsection{Supervised-Learning Classifiers}
Our objective is to efficiently map the eight-dimensional feature space, used for representing bursts, to interference classes. The target interference technologies (802.15.1, BLE beacons\footnote{Although BLE is part of the 802.15.1 standard since release 4.0, we target its identification separately as: 1) it introduces observable differences at PHY, such as larger bandwidth, 2) BLE-based applications are becoming increasingly popular.}, 802.15.4 and 802.11) are represented with the elements of the label set $\mathcal{L}_I=\{B,L,Z,W\}$. In practice, it is a single-label multi-class classification problem and we employ widely known \cite{Bishop:2006:PRM:1162264,Random_forest} classification approaches within the SL framework. Finally, the candidate methods are trained using a common dataset $S^{t}=S_B^{t}\cup S_W^{t}\cup S_L^{t}\cup S_Z^{t}$ engineered basing on experimental observations.

\subsubsection{Classification Tree Family}
Classification trees (CT) are a lightweight and human-readable approach to classification, where points in the feature space are assigned to one of the target classes by using a sequence of decisions (splits). The structure of the tree itself can be generated using a multitude of approaches, spanning from heuristic to SL. 

\begin{figure}
\centering
\includegraphics[width=0.34\textwidth]{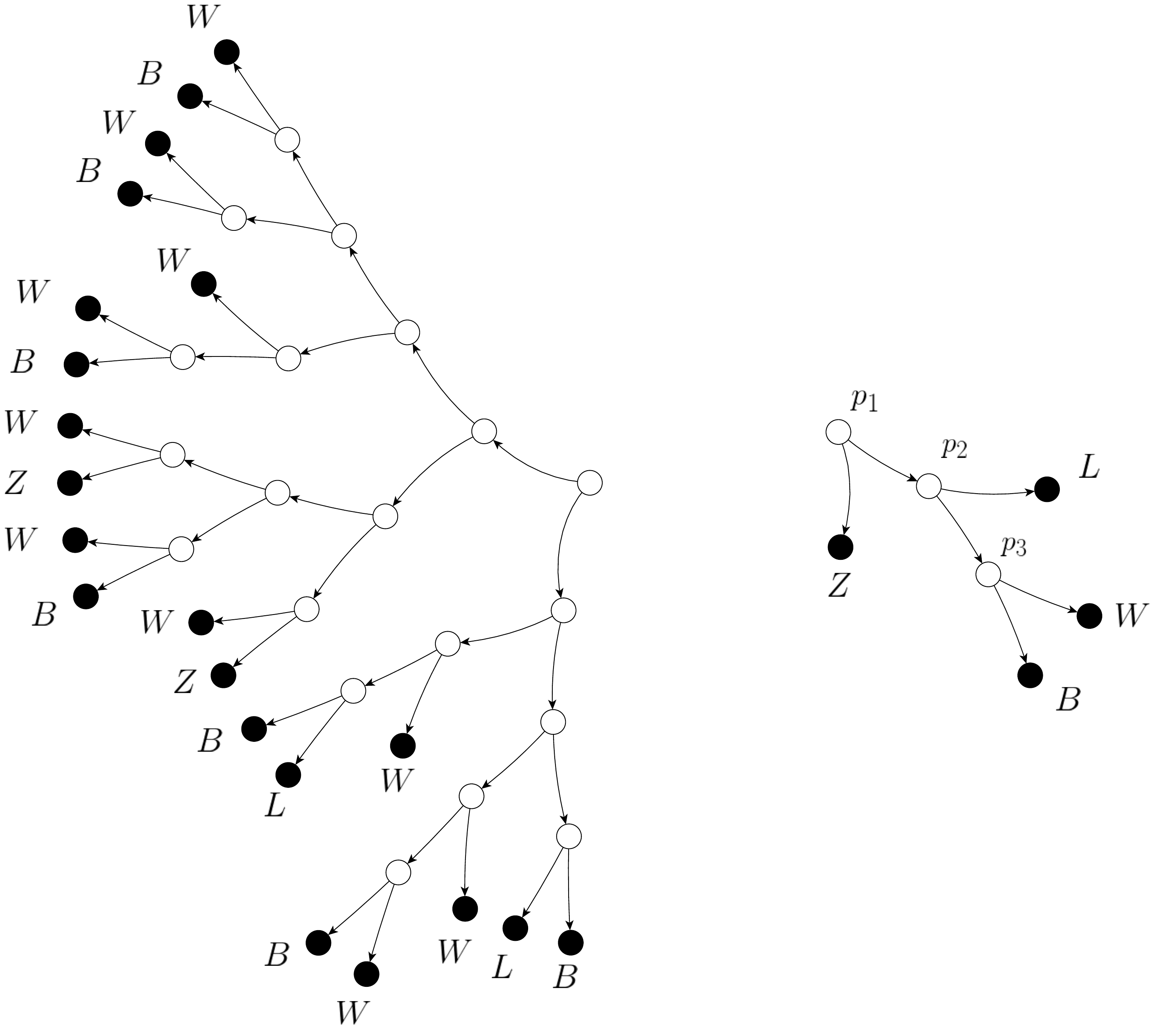}
\caption{From right to left, the custom classification tree CT1 with training-data driven parameter optimization, and classification tree CT2, directly generated with a SL approach.}\label{FIG:classification_trees}
\vspace{-12pt}
\end{figure}

\subsubsection*{Custom CT with SL-driven Parameter Optimization (CT1)}
\begin{figure}
\centering
\includegraphics[width=0.48\textwidth]{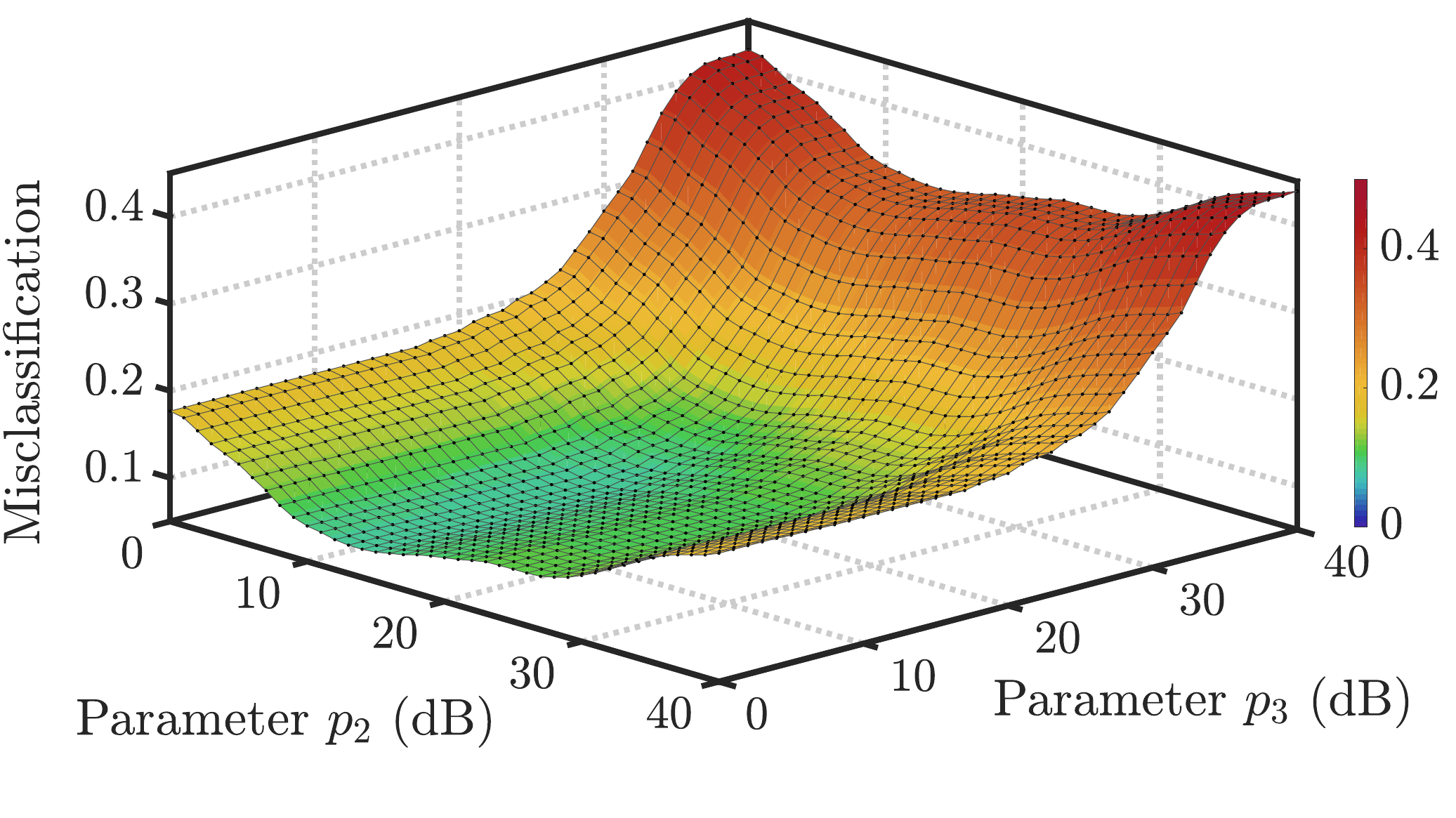}
\caption{SL-based optimization of parameters for CT1: misclassification function $g_m(S^{t},\bar{p})$ in a feasible region of the $(p_2,p_3)$-parameter subspace.}
\label{FIG:Decision_tree_min}
\vspace{-12pt}
\end{figure}

Our first candidate method uses a custom multivariate-split CT, shown in Fig.~\ref{FIG:classification_trees}, meaning that the tree is generated manually, resembling the algorithm-based methods in the literature. However, to detach from the usual heuristic approach, we employ semi-parametric decisions at each split with SL-backed selection of parameters. To such end, we define a misclassification function $g_m(S^{t},\bar{p})$ as the false positive ratio calculated over all the elements $\bar{s}^{t}_{i} \in S^{t}$ with a priori known label $l_{i}^t$, for a certain choice of parameters $\bar{p}=\left[p_1,p_2,p_3\right]$. Hence an optimal set of parameters arises from the solution of the following minimization problem
\begin{multline}
\begin{aligned}
& \text{minimize:} \quad  g_m(S^{t},\bar{p})=\sum_{l \in \mathcal{L}_I}{\frac{\sum_{i=0}^{|S^{t}_l|}{g_e(g_{c}(\bar{s}^{t}_{i},\bar{p})},l^{t}_{i})}{|\mathcal{L}_I| |S^{t}_l|}}& \\
\end{aligned}
\\ \text{such that:} \quad \vert p_{1,2} \vert \leq \frac{D_r}{2}, \quad \text{and} \quad {0} \geq p_3 \leq  D_r
\end{multline}
with $g_c$ CT1-labeling function, $g_e$ evaluation function assigning a penalty $g_e(l_1,l_2)=1$ if the labels $l_1,l_2$ are different and {0} otherwise, while the dynamic range $D_r$ bounds the optimization problem within a feasible range of RSSI, according to the employed hardware.
Specifically, the parameters $p_1$ and $p_2$ are used for thresholding the SF $F_{S_u}$, while $p_3$ is employed in the inequality $\vert F_{S_u}-F_{S_d}\vert>p_3$. The inequalities are then employed in more complex multi-feature decisions at each split. While the general setup of the tree for CT1 is inferable from Fig.~\ref{FIG:classification_trees}, we omit the complete structure of the splits for space reason. Instead, Fig.~\ref{FIG:Decision_tree_min} shows the shape of the misclassification function in the region of global minimum, derived using a grid-search approach.

\subsubsection*{SL-Classification Tree (CT2)} 
In CT1, we developed a customized CT with a simple and intuitive structure that benefits from SL for parameter optimization. On the other hand, in CT2, we employ SL at the very beginning of problem formulation. Therein, the training set $S^{t}$ is used to drive the learning process of a $n_c$-univariate-split decision tree, meaning that at each split a binary decision is made by thresholding a certain feature. The SL process, in this case, can be driven by different minimization targets, e.g., entropy gain, while the number of splits, directly affecting the complexity of the tree, is selected by the parameter $n_c$. Fig.~\ref{FIG:classification_trees} shows the structure of CT2 generated with Gini's diversity index minimization, with a constraint of $n_c \leq {20}$ in order to favor implementability in COTS hardware.

\subsubsection*{Random Forest of Classification Trees (RFCT)}
The complexity and the performance of supervised CT methods depends greatly on the parameter $n_c$. Anyway, extreme values of $n_c$ do not bring classification improvement instead lead to issues of over-fitting and higher complexity. The idea of random forest of classification trees (RFCT) is to generate a set of CTs with sufficient cardinality, pursuing reduction of classification variance \cite{Random_forest}. In our method, we test different sizes of RFCT composed of fully grown SL-CT, i.e., without the constraint $n_c\leq {20}$ as in CT2, to maximize classification accuracy at the cost of increased complexity.

\subsubsection{Multiclass-SVM (MSVM)}
Support vector machine (SVM) is a powerful binary classifier, exploiting quadratic-programming (QP) methods to determine an optimal decision hyperplane in the feature-space. The possibility to use different kernel function ensures good classification performance also in non-linearly separable data sets. In this paper, we use one of the many possible multi-class extensions of SVM: multiple binary SVMs which are trained autonomously using the error-correcting output codes approach. A Gaussian kernel is employed for the single SVM for its proven effectiveness in dual-class signal classification problems~\cite{Grimaldi}.

\subsection{Key Factors for Feasibility}
\label{SUBSEC:Configiration}
Burst-based IDI is at the limit of the capability of COTS hardware, thus proper setup of the radio front-end is strongly recommended. We achieve rapid ($<400 \mskip3mu\mu$s) (de)activation of a \emph{IDI-mode} at run-time such that the normal network operation remains unaffected, while enabling the following features.


\subsubsection{Fast Frequency-Switching}
The frequency-switching time $\Delta T_{sw}$ given in Table~\ref{TAB:sampling_parameters} is considerably smaller than the one reported in the literature (e.g., \cite{Jamlab}), which is often source of erroneous interpretation in spectrum sensing works \cite{ansari_wispot}. As a matter of fact, the 802.15.4 standard mandates firm tolerance on frequency accuracy, i.e. $\pm$40$\mskip3mu$ppm, which reflects in adequate settling time (i.e., 294$\mskip3mu\mu$s for CC2420 \cite{Jamlab}) for the on-board VCO. This constraint can be safely ignored while sampling SFs, as there is no signal demodulation involved, and the reduced VCO accuracy has negligible effect on SFs' extraction. 

\subsubsection{Narrower Frequency Response}
\begin{figure}
\centering
\includegraphics[width=0.47\textwidth,clip, trim=0cm 0cm 0cm 0cm]{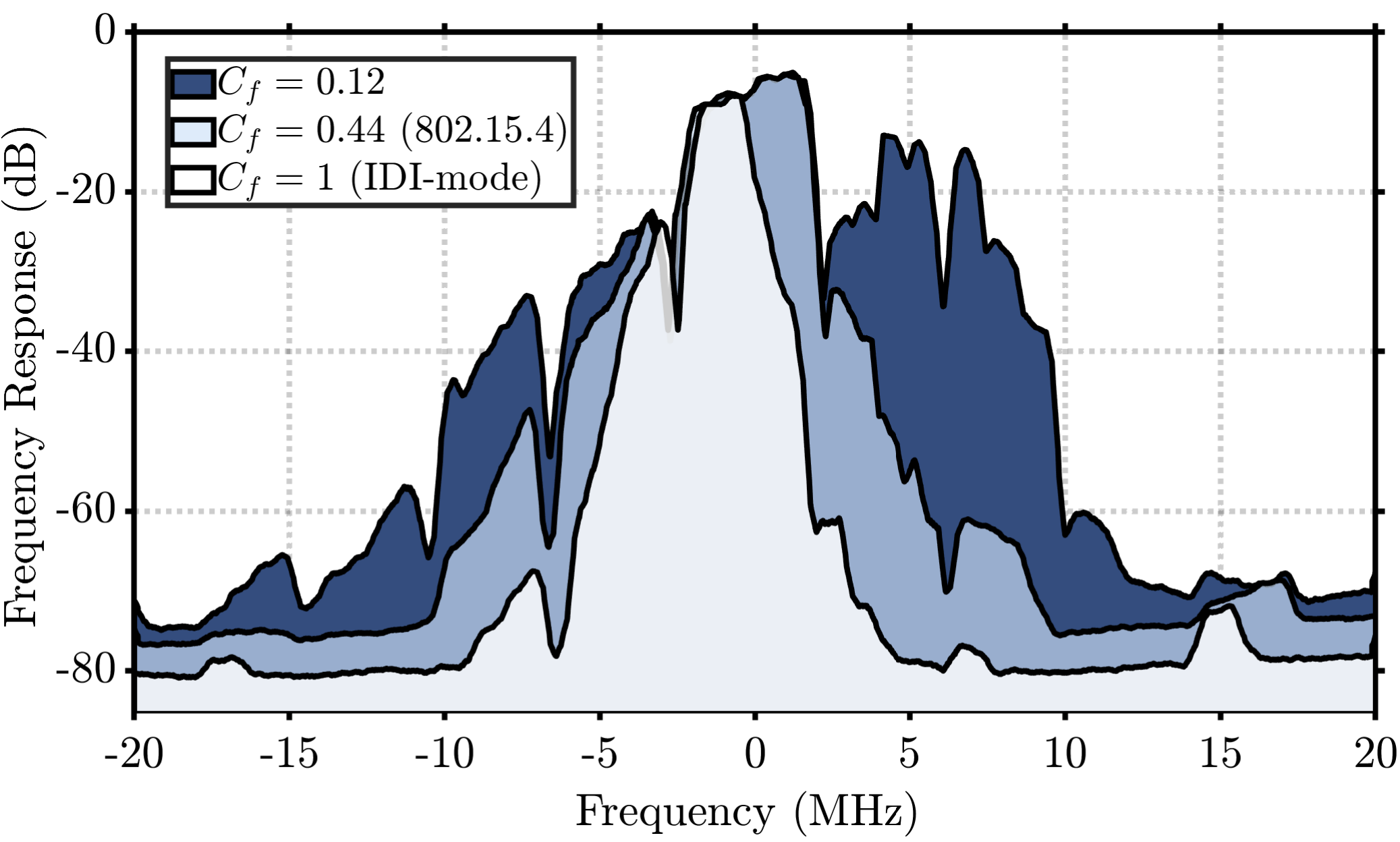}
\caption{Frequency response of CC2420 baseband BPF for different values of normalized calibration capacitance $C_f$. }
\label{FIG:BPF}
\vspace{-13pt}
\end{figure}

The extraction of SFs is in fact a method for spectrum analysis, hence it benefits from higher frequency resolution, especially when targeting narrow-band (i.e., 1$\mskip3mu$MHz-wide) transmissions, such as 802.15.1. The 802.15.4 radios commonly perform channel selection via band-pass-filter (BPF) in digital-domain, while in the CC2420 platform the bandwidth of the BPF is also adjustable (see Fig.~\ref{FIG:BPF}), allowing to narrow the frequency response and therefore to improve the frequency selectivity of the radio.


\subsubsection{RSSI Data Integrity}
According to \cite{Jamlab}, the automatic gain control (AGC) employed in CC2420 receiver chain causes sporadic saturation of the on-board analog-to-digital converter (ADC). This reflects in incorrect RSSI readings, which potentially jeopardize the integrity of all the extracted features. We avoid this by using a solution similar to \cite{Jamlab} for overriding AGC in the amplifier stage.

\section{Experimental Setup}
\label{SEC:Experimental_setup}
\subsection{Hardware Implementation}
We implemented the proposed IDI solution in Contiki OS 3.0 port for Crossbow TelosB WSN motes, which are based on TI CC2420 \cite{CC2420} radio and MSP430 microcontroller. Note that the employed WSN platform is relatively old; however, this selection is deliberate to ensure portability of the solution to wide range of WSN hardware. We also implanted CT1 and CT2 classification variants for online classification. In order to log the online classification results and the raw feature vectors for testing offline reference method (i.e., MSVM and RFCT), a serial interface is utilized. Due to resource-constraints, MSVM and RFCT are implemented in a dedicated laptop using Mathworks MATLAB libraries. In addition, National Instruments USRP-2932 SDR~\cite{USRP2932} and Metageek Wi-Spy spectrum analyzer\footnote{\url{https://www.metageek.com/products/wi-spy/}} are used for validating the experimental setup and for collecting support data. In some experiments, we used Wireshark\footnote{\url{https://www.wireshark.org/}} together with Intel AC7260 802.11 interface to find reference IRN traffic statistics.
 \subsection{Experimental Locations}
We collected the experimental data in four different environments. The description of each follows:
 
 \begin{itemize}
 \item \emph{Location A} is an underground tunnel with no detectable wireless interference. This controlled environment is exploited in all the experiments requiring isolation of the studied interference sources. 
  \item \emph{Location B} is an office area with partly controlled interference due to resident 802.11 IRN and a 30$\mskip3mu$MHz free portion of the spectrum, employed in experiments where the IDI is performed on a limited number of channels.
  \item \emph{Location I1} is a 15$\mskip3mu$m x 25$\mskip3mu$m industrial warehouse in L'Aquila (Italy) heavily cluttered with lathe machines, with a resident 802.11n network operating with hybrid 20/40$\mskip3mu$MHz channel-width, overlapping with $c_Z\in[16,23]$\footnote{From now on we define $c_l$ as the generic channel for the interference technology with label $l\in\mathcal{L}_I$. We refer to Table~\ref{TAB:Technologies_coex} for channel numbering.}.
  \item \emph{Location I2} is the 15$\mskip3mu$m x 15$\mskip3mu$m multi-room workshop with both production and office areas in Sundsvall (Sweden)~\cite{Grimaldi}, with multiple 802.11 IRNs on $c_W=\{1,6,11\}$.
 \end{itemize}
\subsection{Design of Experiments}
The proposed solution is tested under controlled and uncontrolled interference from real hardware. Specifically, we generate controlled interference in locations A and B for assessing the IDI performance with respect to the single interference labels (experiments E1, E2, E4). Conversely, locations with uncontrolled wireless sources are exploited to test the overall capability of the IDI method in real-world scenarios (experiment E3). 
\subsubsection*{E1 - Interference Specific Identification Accuracy}
A WSN node executes the proposed solution at locations A and B and sequentially scans all the $c_Z$ channels, allowing for frequency-domain performance assessment. We ensure that the interference from all the target interference technologies includes 1) wide range of traffic patterns 2) several COTS devices/radios with different variants of standards 3) clear line-of-sight (LoS) path between the WSN node and the interfering device. Distances in the range 0.5$\mskip3mu$m--2$\mskip3mu$m are selected to ensure the collection of bursts with a wide range of INR.

\subsubsection*{E2 - Effects of Spatial Separation and Obstructions}
We investigate whether and how spatial separation and LoS/NLoS affects identification accuracy via specific experiments in Location A. The examined distances span in the range of 0.3$\mskip3mu$m--15$\mskip3mu$m, depending on the detection capability of the devices. 
\subsubsection*{E3 - Real-world Experiments} Preliminary investigations and measurements at locations I1 and I2 showed that interference was exclusively due to 802.11 IRN, ensuring Internet connectivity to personnel. The node with IDI operated in multiple points of I1 and I2 for an overall time of 3$\mskip3mu$h during the production hours, in order to ensure variability of interfering traffic. 
\subsubsection*{E4 - Concurrent Interference} Controlled interference from multiple sources is generated at Location B, in order to test the capability of the IDI to isolate and extract label-specific traffic distribution. 
 
\subsection{Size of Datasets}
In E1 and E2, the IDI system detected and classified over 84000 interference bursts of known origin. The experimental campaign E3 in industrial environments enriched the data set with over 2$\mskip3mu$h of observations yielding about 20000 interference bursts. Finally, the E4 experiments have led to over 9000 bursts, while the labeled dataset $S^{t}$, used for training the SL-classifiers, is engineered with approximately 5000 bursts, distributed unevenly among the interference labels.

\section{Results}
\label{SEC:Results}
\subsection{Interference-specific Identification Accuracy}
\begin{table*}[!h]
\caption{Rate ($\%$) of Predicted Interference Labels and Standard Deviation $\sigma_A$ ($\%$) for Bursts with INR $\geq$ 20$\mskip3mu$dB. The TPR Accuracy is in Boldface.}
\centering
\begin{tabular}{c|c|C{1.7cm} C{1.7cm} C{1.7cm} C{1.7cm}||c} 
\textbf{Method}& \textbf{Interference Source}&\multicolumn{4}{c||}{\textbf{Predicted Label ($\mathbf{\%}$)}}& $\mathbf{\sigma_A ({\%})}$\\
& & \textbf{802.15.1}& \textbf{802.15.1 BLE} & \textbf{802.15.4} & \textbf{802.11}& \\
\hline 
& {802.15.1} & \textbf{{97.30}} & {0}& {0.06} & {2.64}&\\
& {802.15.1 BLE} &  {1.04} & \textbf{{97.41}} &  {0} &  {1.55}&\\
RFCT & {802.15.4} & {0.30}  & {0}  & \textbf{{ 98.92}}  &  {0.78}& \\
{(offline)}& {802.11b} & {4.11}& {0}  & {0.25}  & \textbf{{95.64}}& { 4.97}\\
& {802.11g}	&  {3.28} & {0}  & {0.15}& \textbf{{96.57} }& \\
& {802.11n ({20}{~MHz})} & {2.05} & {0}  & {0.53 }& \textbf{{97.41}}&\\
& {802.11n ({40}{~MHz})} &  {3.47 }& {0}   & {0.47}& \textbf{{96.06}}&\\
\hline
& {802.15.1} & \textbf{{96.62}}  & {0} & {0.26}& {3.11}&\\
& {802.15.1 BLE} &  {2.07}  & \textbf{{97.41}} & {0}& {0.52}&\\
MSVM & {802.15.4} & {0.62}   & {0}  & \textbf{{96.53}}  & {2.85}& \\
{(offline)}& {802.11b} & {6.68}  & {0}  & {0.25}  & \textbf{{93.06}}& { 6.35}  \\
& {802.11g}	& {5.34} & {0}  & {0.16}  & \textbf{{94.49}}& \\
& {802.11n ({20}{~MHz})} & {3.55}  & {0} & {0.92} & \textbf{{95.52}}&\\
& {802.11n ({40}{~MHz})} &{3.90 }  &  {0}   & {0.75} & \textbf{{95.35} }&\\
\hline
& {802.15.1} & \textbf{{90.88}}  & {0}  & {4.85 } & {4.25}&\\
& {802.15.1 BLE} & {18.13 } & \textbf{{79.79}} & {2.07}   & {0}& \\
CT2 & {802.15.4 } & {0.59}   & {0}  & \textbf{{97.72}}  & {1.68} &\\
{(online)}& {802.11b} & {3.12 } & {0.06}  & {3.97} & \textbf{{92.83}} & { 8.26 } \\
& {802.11g}	& {3.85}  & {0.02}& {7.0}  & \textbf{{89.11}} &\\
& {802.11n ({20}{~MHz})}	& {1.87} & {0.04}& {2.35} & \textbf{ {95.72} }&\\
& {802.11n ({40}{~MHz})} &{3.42 }&{ 0.08}   &  {3.19 } & \textbf{{93.29}}&\\
\hline
& {802.15.1} & \textbf{{86.35} }  & {3.28}   & {0}  & {10.35}&\\
& {802.15.1 BLE} & {20.21}& \textbf{{79.79}} & {0}  & {0}& \\
CT1 & {802.15.4 } & {1.91}    & {0.09} & \textbf{{97.86}}  & {0.13} &\\
{(online)}& {802.11b} &  {9.8}   & {0}  & {0.44} & \textbf{{89.74}}&  { 12.96 }\\
& {802.11g}	& {14.66}  & {0} & {0.14} & \textbf{{85.19} } &\\
& {802.11n ({20}{~MHz})}	& {7.79} & {0} & {0.26} & \textbf{{91.93}}&\\
& {802.11n ({40}{~MHz})} & {9.08} & {0}  & {1.01}  & \textbf{{89.91}} & 
\label{TAB: accuracy_global}
\end{tabular}
\end{table*}

In this section, we present the results of experiment E1, where the accuracy of each identification scheme for single interference sources is evaluated in terms of true positive ratio (TPR). TPR is the ratio between the number of correctly identified bursts and the number of detected bursts. A summary of the results is given in Table \ref{TAB: accuracy_global}. In general, we observe that the higher complexity of MSVM and RFCT provides a certain performance edge over CT methods, while CT2 performs consistently better than CT1 for its deeper classification tree. Nevertheless, the elementary decision strategy of CT1 provides a good benchmark on the goodness of the feature selection, reflecting on the degree of separation among interference classes already in the feature space.

All the methods show good accuracy in identifying 802.11 interference, with average TPR of 89.20$\mskip3mu\%$ (CT1), 92.74$\mskip3mu\%$ (CT2), 94.60$\mskip3mu\%$ (MSVM) and 96.42$\mskip3mu\%$ (RFCT). The offline classifiers also ensure a remarkably limited variance over different 802.11 variants. The classification accuracy of 802.15.4 transmissions is exceptional as well, where all the methods maintain the average TPR of $\geq 96\mskip3mu\%$. In this case, even CT1 performs better than MSVM, because the $F_{\textrm{CCA}}$ feature enables strong separation in the feature space (i.e., 97.86$\mskip3mu\%$ of detected 802.15.4 bursts resulted in a negative CCA), which limits the advantage of more elaborate classification methods. 

When it comes to discerning BLE from 802.15.1 signals, there is a noticeable difference between the offline and online methods. While offline RFCT and MSVM misclassify only 2$\mskip3mu\%$ or less of BLE bursts as traditional 802.15.1, the misclassification is 10 times higher for online methods. This behavior is caused by the strong similarity between the two technologies at the PHY layer (i.e., both utilize GFSK modulation), which leads to very similar spectral footprint. This, in turn, produces subtle differences at the SF-level, and thus mandates specialized classification methods.
The identification of 802.15.1 bursts follows a similar trend, where the TPR gap between the best online and offline methods is 6$\mskip3mu\%$. Interestingly, in RFCT the rate of false negatives between 802.11 and 802.15.1 signals is 90$\mskip3mu\%$ and 97$\mskip3mu\%$ respectively, meaning that more than 9 out of 10 of 802.15.1 misclassified bursts are confused as 802.11 and vice versa. The cause of this behaviour is elaborated in Section \ref{SEC:Discussion}. 

\begin{figure}[!h]
\centering
\includegraphics[width=0.48\textwidth,clip, trim=0cm 0cm 0cm 0cm]{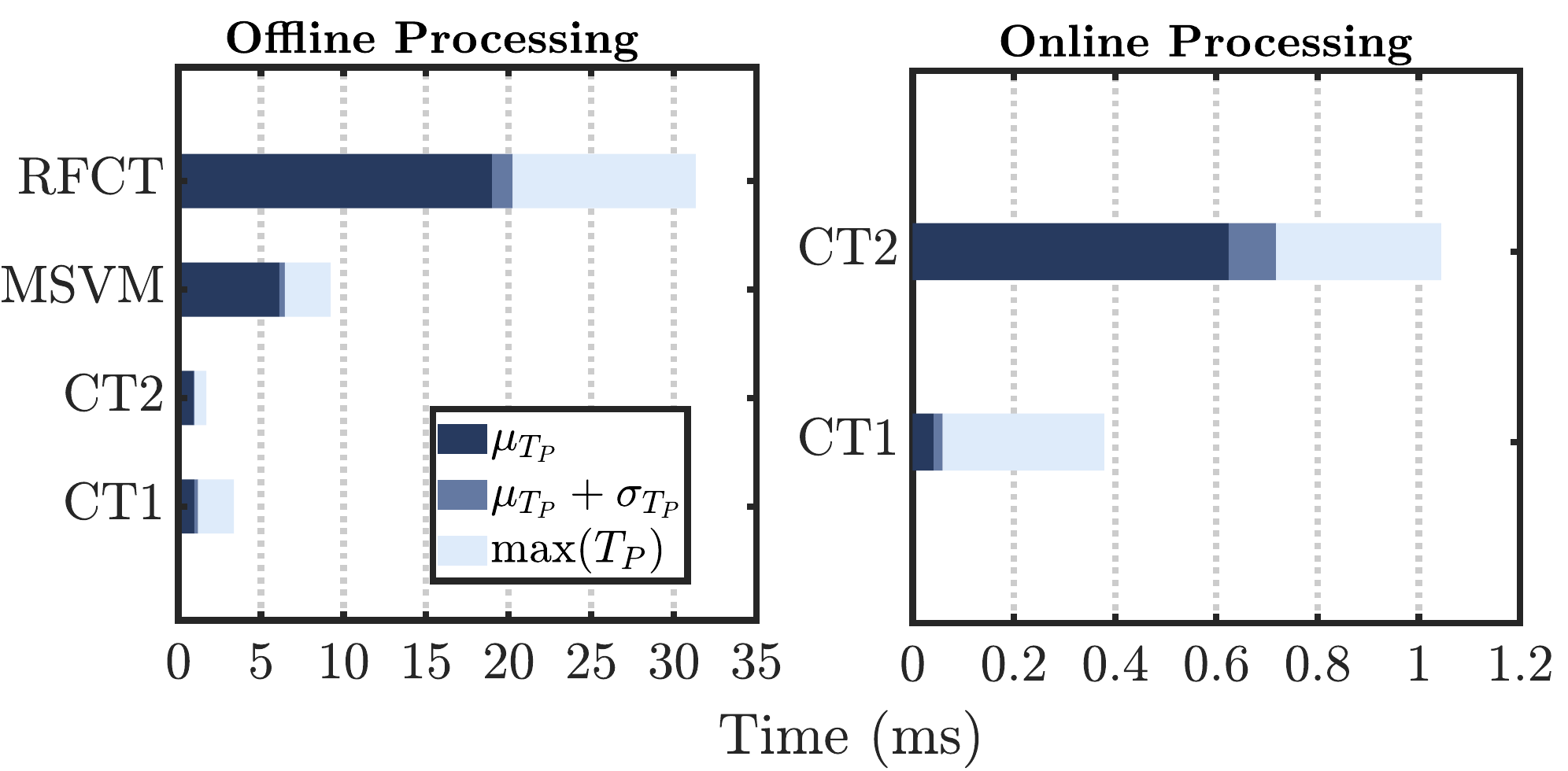}
\caption{Execution time ($T_P$) of the proposed burst identification methods for offline and run-time (TelosB) implementations.}
\label{FIG:processing_delay}  
\vspace{-12pt}
\end{figure}

In Fig.~\ref{FIG:processing_delay}, we show the execution time of the classification methods in their online and offline implementation in TelosB and MATLAB, respectively. Both CT1 and CT2 benefit from a simple structure and an optimized implementation, showing sub-ms execution times and approximately 1$\mskip3mu$ms worst-case delay. Considering the high accuracy of CT2 in identifying each interference burst with real-time constraints, this is a promising outcome. Whereas, the slightly better accuracy of MSVM and RFCT comes at a cost of significantly higher processing time with respect to offline CT methods.


\subsection{Full-spectrum Performance}
\begin{figure*}[t]
\centering
\includegraphics[width=1\textwidth,clip, trim=0cm 0cm 0cm 0cm]{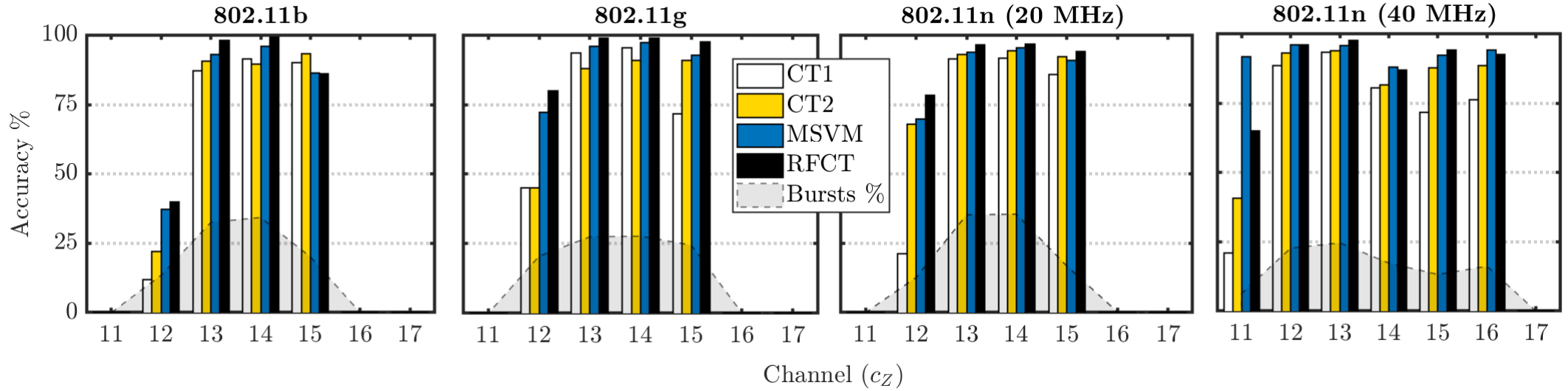}
\caption{Per-channel identification accuracy and ratio of identified bursts for 802.11 interference on channels $c_W=2$ during experiment E1.}
\label{FIG:result_channelspec_wifi}
\end{figure*}

We analyze the effect of the relative frequency offset between $c_Z$ and $c_W$ on the identification of 802.11 bursts, using the experimental methodology E1. Fig.~\ref{FIG:result_channelspec_wifi} shows the channel-specific TPR for different versions of the 802.11 standard. It can be observed that when $c_Z$ overlaps with the central region of $c_W$ the TPR is consistently high, whereas a drops occurs on the leftmost $c_Z$, suggesting a possible lack of separation in the SF sub-space. Furthermore, all the methods show rather poor performance for 802.11b interference on this channel, whereas MSVM and RFCT ensure some performance gain for the g/n versions of the standard.
We explain the behavior noting that the 802.11b and 802.11g/n standards implement different PHY, leading to spectral masks with relevant differences around the $\pm$10$\mskip3mu$MHz regions from the central frequency~\cite{80211_mask}---it perfectly matches our observations, considering the 5$\mskip3mu$MHz-width of $c_Z$. 
However, due to spectral shape of 802.11 signals, the percentage of bursts captured on this side-channel is limited, meaning that the impact of this local drop of overall identification performance is minimum.

\begin{figure*}[t]
\centering
\includegraphics[width=1\textwidth,clip, trim=0cm 0cm 0cm 0cm]{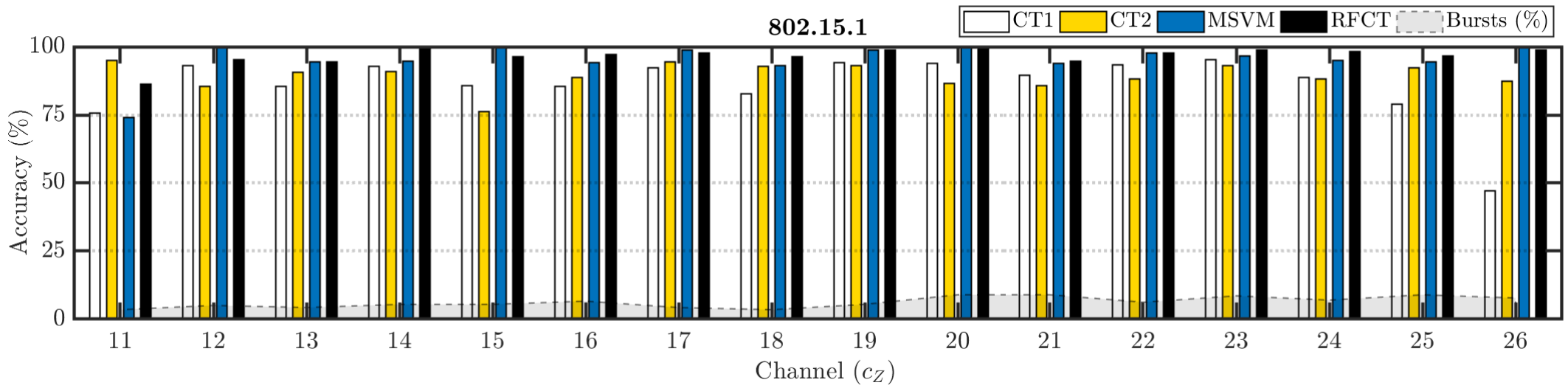}
\caption{Per-channel identification accuracy and ratio of identified bursts for 802.15.1 interference during experiment E1.}
\label{FIG:result_channelspec_bt}
\vspace{-12pt}
\end{figure*}

We repeat the analysis for 802.15.1 interference, and perform IDI on all available $c_Z$ channels. The TPR shown in Fig.~\ref{FIG:result_channelspec_bt} is generally consistent across the spectrum, except CT1 that suffers performance inconsistency on the extreme sides of the ISM band. In addition, we observe a slight performance drop on $c_Z=\{11,15,26\}$, likely due to the fact that these channels are overlapped by BLE broadcast channels $c_L=\{37,38,39\}$, which are used in BLE beacon applications. CT2, MSVM, and RFCT show higher capability to cope with the additional BLE class on these channels, minimizing false negative rate and reflecting in overall better performance.

In Table \ref{TAB: accuracy_global}, we include a stability measure of the identification methods in frequency domain for 802.11 and 802.15.1 interference, by means of the full-spectrum standard deviation for channel classification accuracy  $\sigma_A$($\mskip3mu\%$). Unsurprisingly, there is only a limited 3$\mskip3mu\%$ gap among the first three classification approaches, highlighting the solid full-spectrum performance for CT2, whereas CT1 shows worse performance.

\subsection{Performance in Real Environments}
\begin{table}
\caption{Experiments E3 in Real Environments: Average Accuracy for Identification of 802.11x bursts.}
\centering
\begin{tabular}{c|C{0.8cm} C{0.8cm}  C{0.8cm}  C{0.8cm} }
\textbf{Location}/&\multicolumn{4}{c}{\textbf{Accuracy ($\mathbf{\%}$)}} \\
\textbf{Position}&\textbf{RFCT}&\textbf{MSVM}&\textbf{CT2}& \textbf{CT1}\\
\hline 
I1/A& {84.88} & {87.51} &{92.40} &{65.91} \\
I1/B& {98.39} & {99.06} & {98.19} &{96.31}\\
I1/C& {98.05} & {91.27} & {96.32} & {82.90}\\
I1/D& {99.87}  & {99.61}& {99.74}& {99.10}\\
\hline 
\textbf{I1/Global}&\textbf{{92.57}} &  \textbf{{93.08}}& \textbf{{95.86}}& \textbf{{82.46}}\\
\hline \hline
I2/A&{91.57} & {91.57}  &{87.95} &{71.08}\\
I2/B&{100}&  {99.70}&{99.90}&{99.31}\\
I2/C&{96.27} & {91.19}& {90.57} & {80.50}\\
I2/D& {99.35} & {97.97}& {98.71}& {97.97}\\
\hline
\textbf{I2/Global}&\textbf{{99.14}} &  \textbf{{98.03}}& \textbf{{98.29}}& \textbf{{96.41}}\\
\hline
\end{tabular}
\label{TAB:table_industrial}
\vspace{-12pt}
\end{table}
In Table \ref{TAB:table_industrial}, we show the results of experiments E3 in two heterogeneous office/industrial environments. In each environment, we collected the data traces at four different locations. Due to the presence of uncontrolled 802.11 interference only, the IDI performance is mainly influenced by the ratio between the burst detected in the central part of the band and on the leftmost in-band-$c_Z$, as previously analyzed. 
All IDI variants show performance comparable to what we achieve in controlled environment while CT2 matches, even improves in some cases, the accuracy of online identification. 
This is mainly explained with the fact that CT2 performs remarkably good with 802.11n interference, which was the predominant choice for the networks found at the experimental sites.

\subsection{Influence of Distance and Interference-to-Noise Ratio}
\begin{figure}
\centering
\includegraphics[width=0.48\textwidth,clip, trim=0cm 0cm 0cm 0cm]{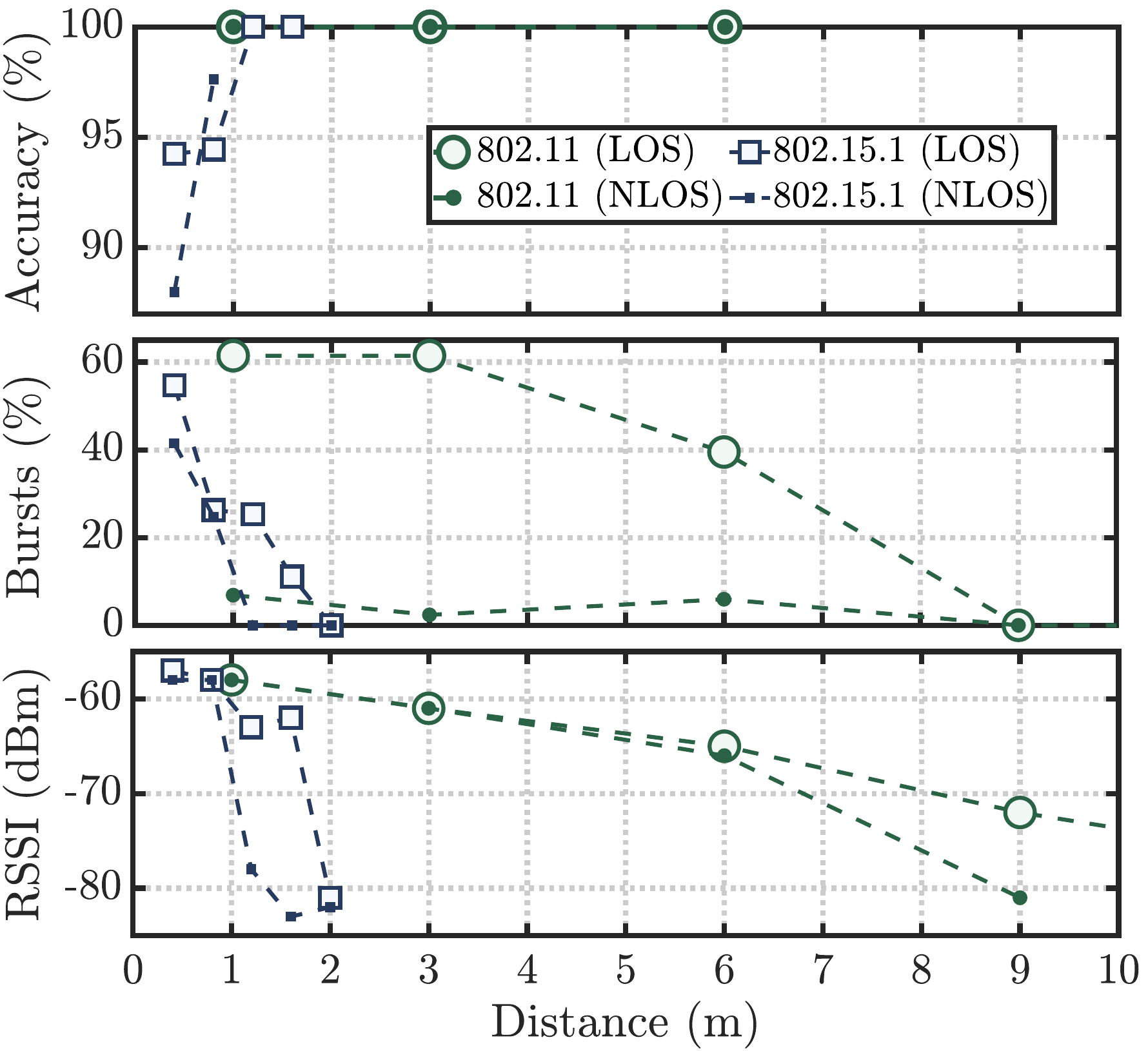}
\caption{Classification accuracy for CT2 with respect to the distance and the LoS/NLoS between source of interference and WSN node. The percentage and the mean RSSI of identified bursts is shown.}
\label{FIG:distance}
\vspace{-12pt}
\end{figure}
We also evaluate the impact of distance between the interference source and the sensing node in LoS/NLoS conditions on the classification accuracy. The experiments, repeated for 802.11 and 802.15.1 interference, show that (see Fig.~\ref{FIG:distance}) the CT2 classification performance does not deteriorate with the distance, whereas the shape of the extracted features is also unaffected by the LoS/NLoS conditions. The result indicates that CT2 classifier can nicely compensate for the influence of multipath-fading on the extracted features. The apparent increase in its accuracy is due to the reduced influence of the side parts of the spectrum, that is, the burst samples from the side parts fall below the selected noise threshold. 

\begin{figure}
\centering
\includegraphics[width=0.48\textwidth,clip, trim=0cm 0cm 0cm 0cm]{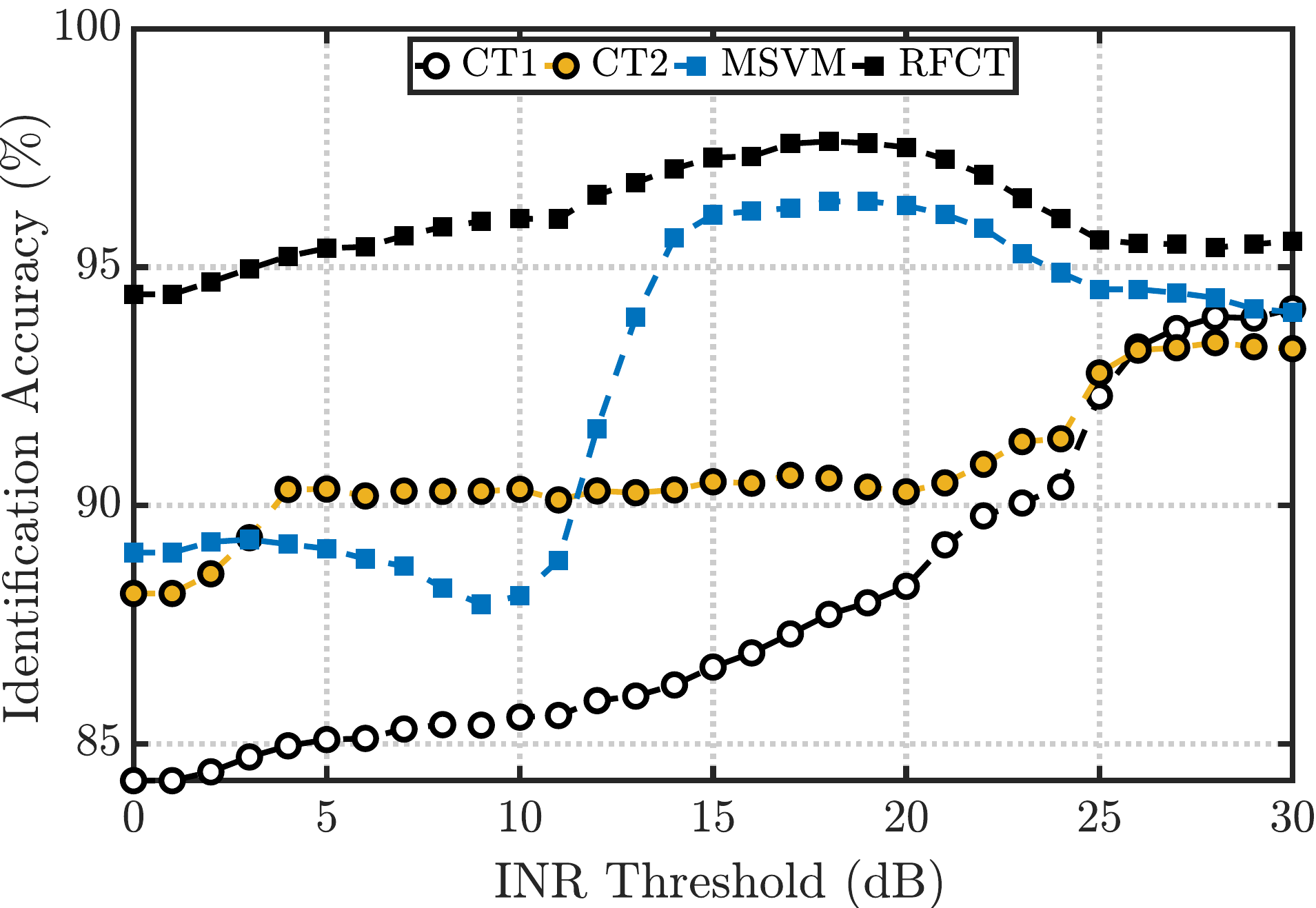}
\vspace{-7pt}
\caption{Identification accuracy of the different supervised classification methods, for various minimum allowed INR.}
\label{FIG:classifiers_INR_results}
\vspace{-12pt}
\end{figure}

The selection of the threshold therefore plays a significant role in the trade-off between the sensitivity and the accuracy of the IDI system. Intuitively, the bursts with lower INR reduce their separation in the feature space, and thus reduce the identification accuracy. This trade-off is analyzed in Fig.~\ref{FIG:classifiers_INR_results}, which shows how different methods respond with respect to INR threshold $\gamma_T$ in terms of identification accuracy. These results are obtained by averaging the identification accuracy results for all the interference classes.
The results show that the RFCT is a good choice for any value of $\gamma_T$, while CT2 performs better than MSVM if $\gamma_T \leq 12 \mskip3mu$dB. In general, the gap among classifiers reduces with the increase in INRs. This result gives important insights on which IDI variant to use, or which threshold to employ for a target identification performance. In the next section, we analyze the physical reasons of the dependence between TPR and INR, and provide an analytical model to predict the related performance improvement.

\section{Analysis of Spectral Features}
\label{SEC:Discussion}
In this section, we extensively investigate the properties of the SFs, in order to ensure, a) an appropriate (instead of heuristic) selection of the frequency parameters for the extraction of SFs, and b) a rationale behind their impact on IDI by estimating an upper bound on the SF-driven identification accuracy gain. 
The analysis is based on the assumption that the variation of the spectrum within each interference labels $l \in \mathcal{L}_I$ (due to different modulation techniques and data-rates) is less relevant that the variation among the different interference labels. Since the latter is governed by rigid standard-related spectral masks, our analysis only requires knowledge of the spectra of target interfering signals $X_l(f), \, \forall  l\in \mathcal{L}_I$ and the frequency response of the employed radio front-end $H(f)$, together with the basic concepts from signal theory. 

\subsection{Physical Background}
\label{SSEC:Abalysis_of_SF}
The first step is to represent SFs in terms of $X_l(f)$ and $H(f)$. Assuming both $X_l(f)$ and $H(f)$ centered at the same frequency $f_c$, we can write their spectral convolution as
\begin{equation}
Y_{l}(\Delta f)=\int_{- \infty}^{\infty} X_l(f)H(\Delta f-f) df
\label{EQ:spectrum_convolution}
\end{equation}
which represents the total power received by the radio front-end with frequency response $H(f-\Delta f)$ for any possible $\Delta f$-wide shift, in {MHz}, around the central frequency. In this case, the SFs defined in \eqref{EQ:SF_definition} are simply given by $F_{S}=Y_{l}(f_c)-Y_{l}(f_c \pm \Delta f)$, with $X_l(f)$ and $H(f)$ centered at $f_c$ and the SFs calculated at $\pm \Delta f$. However, considering that the channel allocation layout differs among target technologies, the offset ($\delta_f$) between the central frequencies of channel used for sensing ($c_Z$) and the interference channel ($c_l$) must be taken into account. This leads to express the SFs as
\begin{equation}
F_{S}=Y_{l}(f_c+ i \delta_f)-Y_{l}(f_c + i\delta_f \pm \Delta f)
\end{equation}
with $i \in \mathbb{Z}$. Since the minimum spacing between the center frequency of two generic channels $c_l$ is 1$\mskip3mu$MHz, $\delta_f$ is governed by the minimum frequency-step of employed hardware\footnote{This value is commonly 1$\mskip3mu$MHz for 802.15.4 radios (see Table \ref{TAB:sampling_parameters}).}. Similarly, the frequency shift used for SFs' extraction can be selected in a discrete fashion as $\Delta f=j\delta_f$ where $j \in \mathbb{Z}$.
Considering the effect of INR ($\gamma_l$) of $X_l(f)$, we can define an SF-variation function as
\begin{equation}
v^{(l)}_{j,\gamma}(i)=w_{i,\gamma_T}[Y_{l}^{(\gamma_l)}(f_0+i\delta_f)-Y_{l}^{(\gamma_l)}(f_0+(i+j) \delta_f)]
\label{EQ:SF_Features_Final}
\end{equation}
where $w_{i,\gamma_T}$ are weighting coefficients defined as
\begin{align}
w_{i,t}= \left\{ \begin{array}{cc} 1 & \hspace{4mm} \textnormal{if} \hspace{2mm} Y_{l}^{(\gamma_l)}(f_0+i\delta_f)>P_T+\gamma_T   \\
0 & \hspace{4mm} \textnormal{elsewhere} \\
\end{array} \right.
\label{EQ:weighting_coef}
\end{align}
where $\gamma_T$ is the eventual INR-based thresholding employed in the IDI system.

From \eqref{EQ:SF_Features_Final}, we note that for each choice of interference label $l$, SF shift $j$ and INR of burst $\gamma_l$, the function $v^{(l)}_{j,\gamma}(i)$ yields a vector with variable number of elements, by means of the index $i$. We can see each of these vectors as the realization of a random variable (RV) $v^{(l)}_{j,\gamma}$ with mean $\mu_{v^{(l)}_{j,\gamma}}$, variance $\sigma^2_{v^{(l)}_{j,\gamma}}$, but unknown distribution.
\begin{figure}
\centering
\includegraphics[width=0.48\textwidth,clip, trim=0cm 0cm 0cm 0cm]{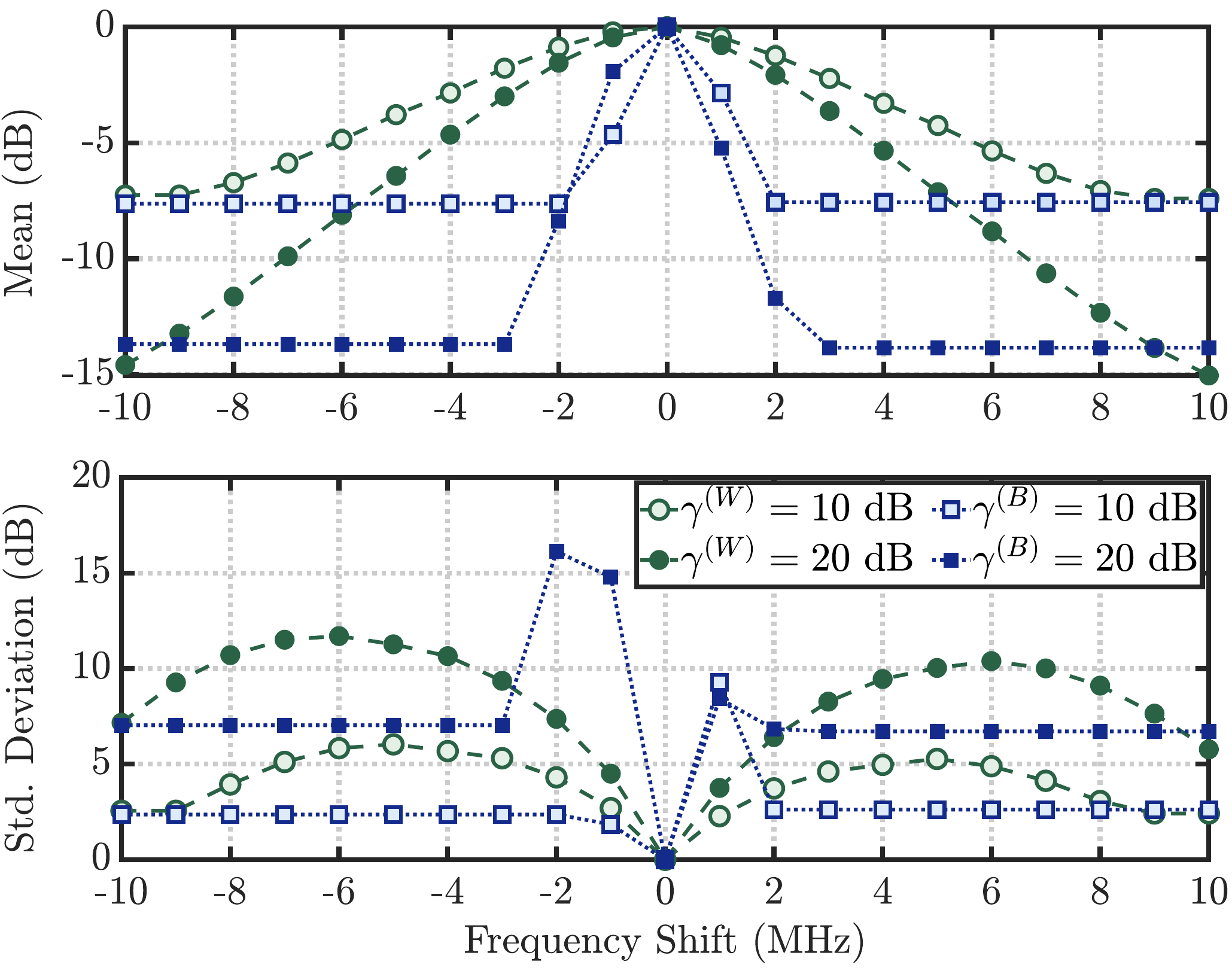}
\caption{First two moments of the SF-variation RV for 802.11 ($l=W$) and 802.15.1 ($l=B$) signals for three distinct INRs $\gamma^{(l)}$.}
\label{FIG:variation_function}
\vspace{-12pt}
\end{figure}

In Fig.~\ref{FIG:variation_function}, we show $\mu_{v^{(l)}_{j,\gamma}}$ and $\sigma^2_{v^{(l)}_{j,\gamma}}$ for $B$ and $W$ interference labels and three different INRs $\gamma_l$. For this analysis, we use signal spectra captured with SDR hardware, while $H(f)$ represents the frequency response of CC2420 radio in \emph{IDI-mode}, as shown in Fig.~\ref{FIG:BPF}.
Fig.~\ref{FIG:variation_function} highlights the influence of the employed frequency shift $\Delta f$ on the characteristics of SFs. For instance, we can observe that the asymmetry of $H(f)$ around $f=0$ has limited influence on the RV, especially on its variance. Overall, this analysis gives a first insight on the existence of SF-level differences between the two interference labels, meaning that the introduction of SFs is expected to bring a certain INR-dependent identification gain. 

\subsection{Frequency Shift Selection}
\label{SSEC:Discussion_freq_offset}
In this section, we use the SF-variation model in \eqref{EQ:SF_Features_Final} for optimizing the selection of the frequency shift. First, we strengthen our model to include the effect of bursts with different INR, and introduce an appropriate similarity measure among interference labels that must be minimized. 
For this purpose, finding the exact distribution of the RVs $v^{(l)}_{j,\gamma}$ appears unnecessarily demanding. Instead, we assume that $v_{j,\gamma}$ can be approximated with a Gaussian probability distribution function (PDF) i.e., $v^{(l)}_{j,\gamma} \sim \mathcal{N}(\mu_{v_{j,\gamma}^{(l)}},\sigma_{v_{j,\gamma}^{(l)}}^2)$. Then, we can introduce a similarity measure between any two interference labels $A$ and $B$, with INR $\gamma^{(A)}$ and $\gamma^{(B)}$ respectively, by a SF-similarity function $S_j(\gamma^{(A)},\gamma^{(B)})$, defined as the overlapping area of the PDFs of $v_{j,\gamma}^{(A)}$ and $v_{j,\gamma}^{(B)}$, that is
\begin{equation}
S_j(\gamma^{(A)},\gamma^{(B)})\!=\!1-Q\!\left(\!\frac{x_c-\mu_{v_{j,\gamma}^{(B)}}}{\sigma_{v_{j,\gamma}^{(B)}}}\!\right) + Q\!\left(\!\frac{x_c-\mu_{v_{j,\gamma}^{(A)}}}{\sigma_{v_{j,\gamma}^{(A)}}}\!\right)
\label{EQ:Error_function}
\end{equation}
where $x_c$ is the intersection point of the two PDFs.
The measure $S_j(\gamma^{(A)},\gamma^{(B)})$ represents the similarity of the SFs extracted with frequency offset $\Delta f = j\delta_f$.
Note that the INR of interference bursts can vary unpredictably according to e.g., distance, LoS, transmission power. Therefore, the similarity function must be calculated for each feasible combination of INR of the two interference classes under analysis. Assuming the bursts with uniformly distributed INR in the feasible dynamic range, the definition of SF-error function $E_{j}^{(A,B)}$ follows as 
\begin{equation}
E_{j}^{(A,B)}=\frac{1}{K_A K_B}\sum_{m=0}^{K_A-1} \sum_{n=0}^{K_B-1}  w_{m,n}S_j(\gamma^{(A)}_{m},\gamma^{(B)}_{n})
\label{EQ:ErFunction}
\end{equation}
where the weighting coefficients $w_{m,n}$ are taken with unitary values, based on uniformly distributed INR, and the interference label $l$ is observed with $K_l$-discrete INR values $\gamma^{(l)}_{m}$ in the range $[\gamma^{(l)}_{0}, \gamma^{(l)}_{K_l}]$. The function $E_{j}^{(A,B)} \in [0,1]$ represents the probability of misclassification of the interference labels $A$ and $B$ in the SF-feature space.

\begin{figure}
\centering
\includegraphics[width=0.48\textwidth,clip, trim=0cm 0cm 0cm 0cm]{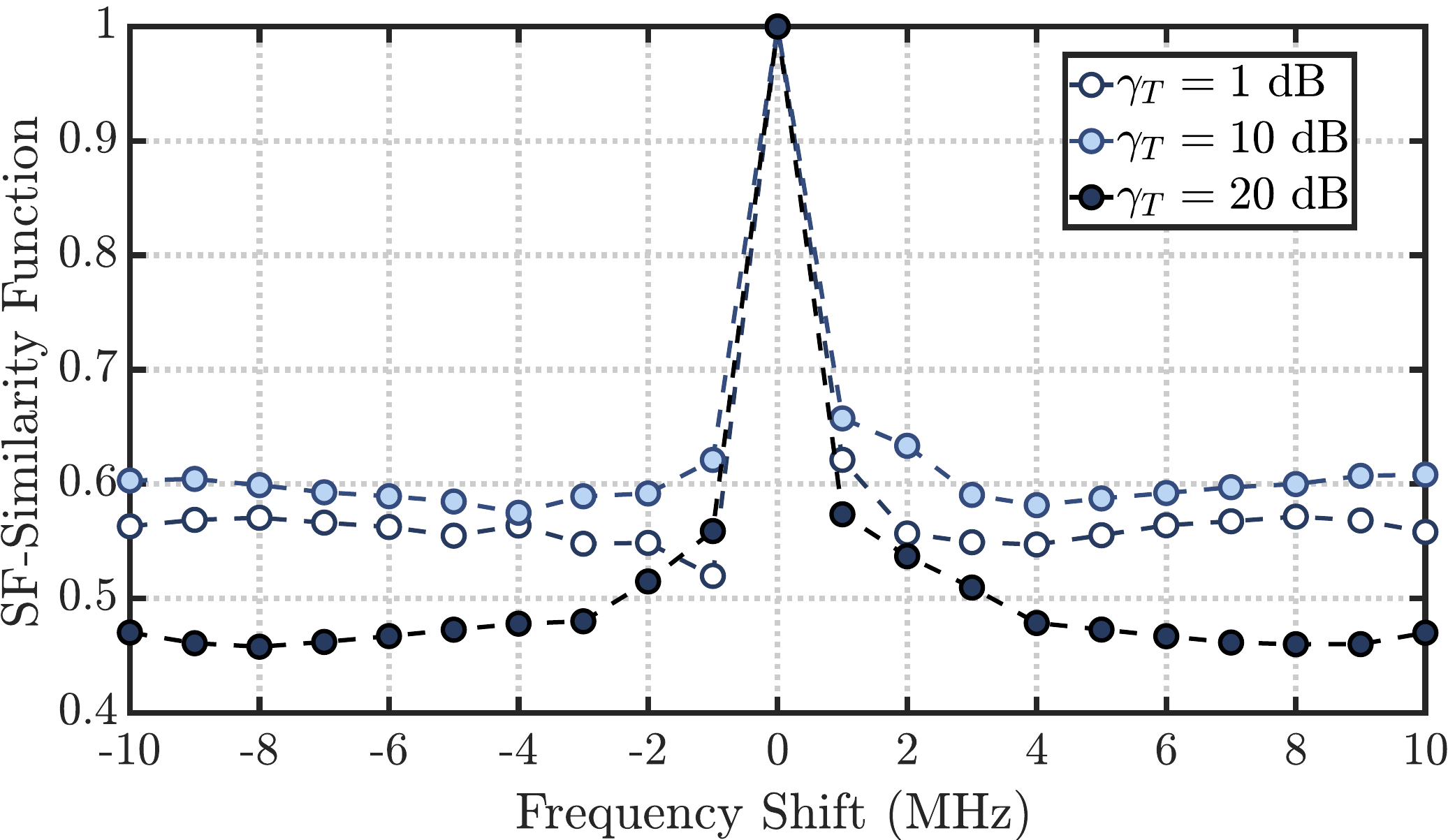}
\caption{SF-similarity function $S_j$ for interference labels $W$ and $B$ computed for different values of frequency shift $\Delta f$.}
\label{FIG:Model_Pe}
\vspace{-12pt}
\end{figure}
In Fig.~\ref{FIG:Model_Pe}, we plot the error function for the labels $W$ and $B$ in the INR range {1}{~dB}--{30}{~dB} for three different INR filtering thresholds $\gamma_T$. This analysis clearly shows the effect of different frequency offsets for SFs' extraction on the classification error. In practice, we observe that $|j \delta_f | \geq 2 \mskip3mu$MHz do not bring significant improvement in terms of reducing error probability, and for $\geq 3 \mskip3mu$MHz the function is almost flat. Also, the higher values of $j \delta_f$ should be avoided since the required central frequency might be out of the capability of the employed hardware\footnote{e.g., CC2420 radio allows only a -5$\mskip3mu$MHz and +3.5$\mskip3mu$MHz excursion over the first and last $c_Z$, respectively, while other hardware, such as TI CC2538 shows wider frequency range.} when scanning the channels at the boundaries of the ISM band (i.e. $c_Z=\{11,26\}$).

We also investigate the effect of increase in $\gamma_T$, which is expected to reduce error probability however at the cost of a loss of sensitivity for low-INR bursts. The analysis shows that for $\gamma_T \leq 15\mskip3mu$dB the filtering causes a slight increase in error probability. Conversely for larger $\gamma_T$ values, the error probability shows a 10$\mskip3mu\%$--15$\mskip3mu\%$ reduction, giving some insights on the impact of INR threshold in real scenarios. We investigate the reason and the impact of this effect in detail in the next Section.

Finally, we acknowledge that a complete analysis of the SF-space would require the calculation of $E_{j}^{(A,B)}, \, \forall (A,B) \in \mathcal{L_I}$, and the solution of the related minimization problem. Anyway, this analysis only aims to understand physically meaningful parameters to employ for SFs' extraction without claims of optimality, therefore leading to the selection of $\Delta{f_u}=\Delta{f_d}= 2 \mskip3mu$MHz.

\subsection{Estimating the Identification Accuracy Gain}
In this section, we estimate an upper bound on the mean classification gain due to the introduction of SFs, in particular for interference labels $W$ and $B$. From the error function in \eqref{EQ:ErFunction}, note that $\bar{E_{j}}^{(A,B)}=1-E_{j}^{(A,B)}$ represents the average probability of correct identification between the generic classes $A$ and $B$. Therefore, we can estimate an upper-bound for the mean identification accuracy (MIA) for a generic classifier $C$ employing only SFs as
\begin{equation}
\hat{A}_C^{(A,B)}\left(\gamma_T\right)=A^{(A,B)}_{C}(0)+\left(1-A^{(A,B)}_{C}(0)\right)\bar{E_{j}}^{(A,B)}\Bigr|_{\substack{\gamma \geq \gamma_T}}
\label{EQ:model_estimator}
\end{equation}
where  $A^{(A,B)}_{C}(0)$ is the MIA of $C$ without INR thresholding (i.e., $\gamma_T=0$) while $\gamma_T$ is used to investigate whether the proposed model can predict the increased identification accuracy of bursts with higher INR observed in Section \ref{SEC:Results}. To validate the model, the employed classifiers are also trained to operate in two additional modes: a) only with SFs and b) without SFs. The mean TPR is then evaluated over the experimental datasets including $B$ and $W$ interference sources, according to the procedure in Section \ref{SEC:Results}.
\begin{figure}
    \centering
    \subfloat[CT2 classifier.]{%
       \includegraphics[width=0.98\linewidth,clip, trim=0cm 0cm 0cm 0cm]{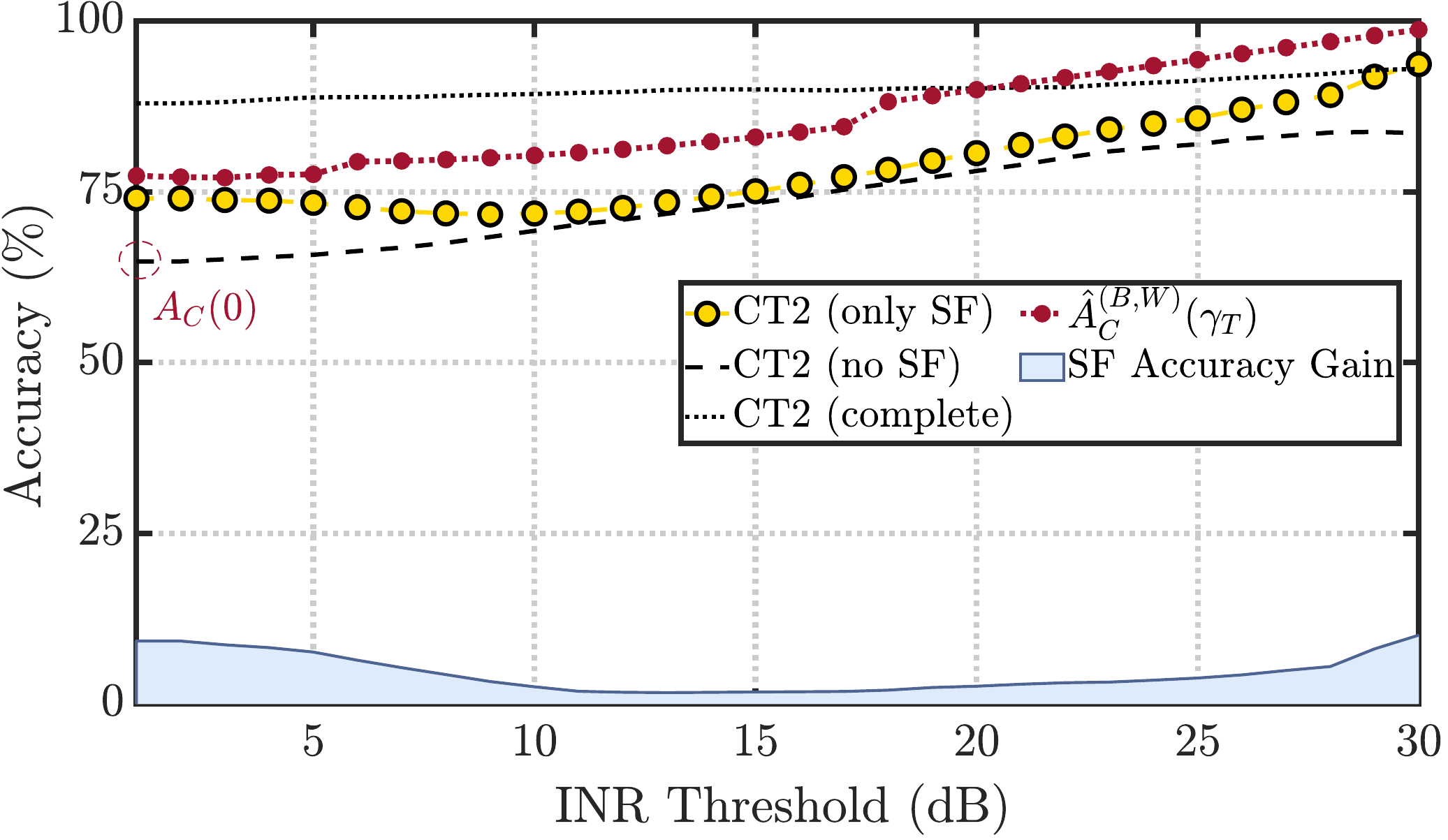} \label{FIG:Model_vs_global_DT}}\hfill
  \subfloat[MSVM classifier.]{%
        \includegraphics[width=0.98\linewidth,clip, trim=0cm 0cm 0cm 0cm]{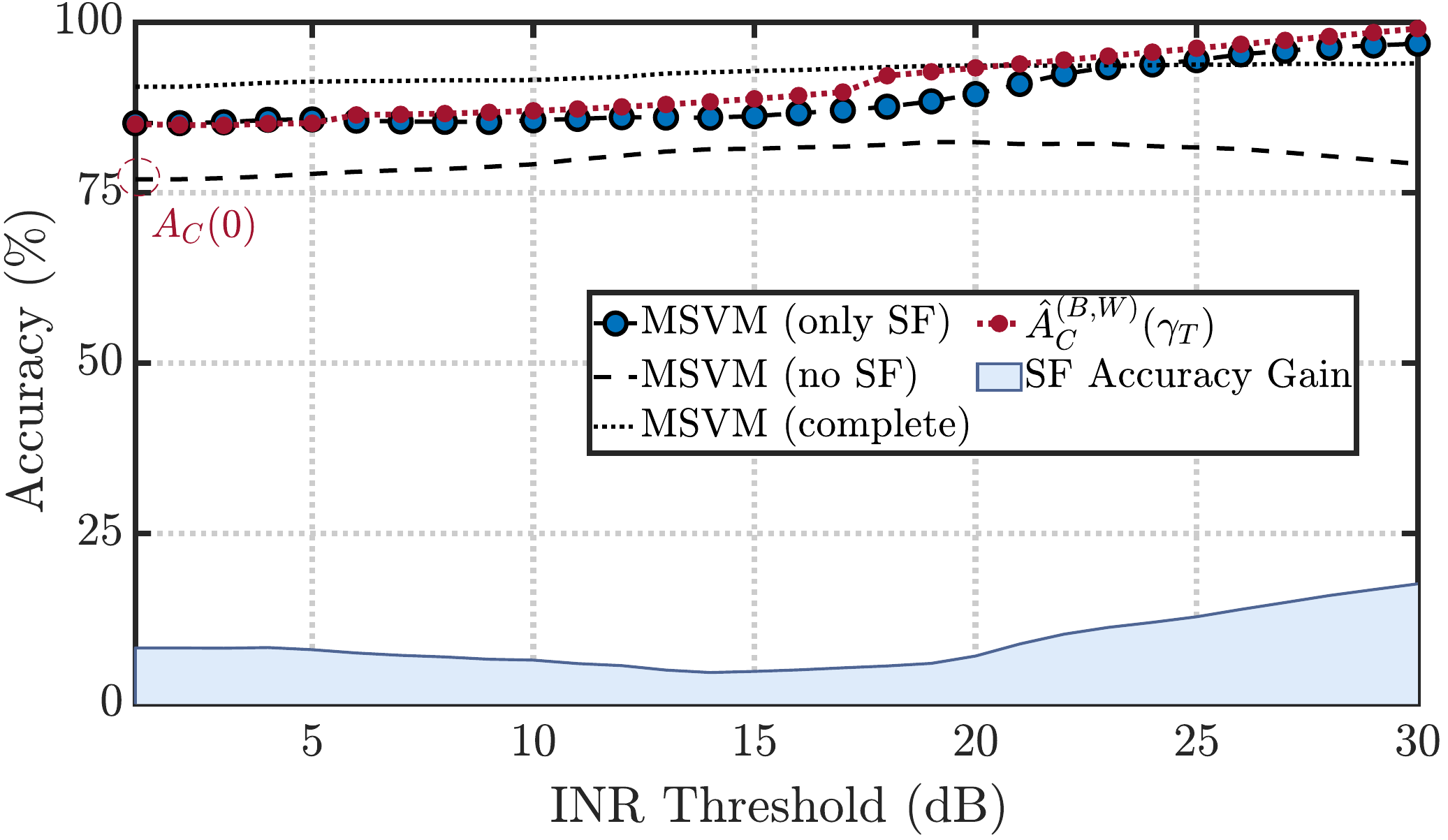} \label{FIG:Model_vs_global_MSVM} }       
\caption{Estimated mean identification accuracy (MIA) upper bound $\hat{A}_C^{(A,B)}(\gamma_T)$ for the interference labels $W$ and $B$ compared with the experimental results from two classification methods.}
 \label{FIG:Model_vs_global} 
\end{figure}

In Fig.~\ref{FIG:Model_vs_global}, we show a comparison between the estimated MIA upper-bound using 
\eqref{EQ:model_estimator} and the experimentally-evaluated MIA for offline MSVM and online CT2 classification methods. The proposed model appears to closely reflect the results in Section~\ref{SEC:Results}, with MSVM ensuring higher MIA then CT2. The gap between upper-bound and experimentally-derived MIA of the classifiers employing only SFs is also tighter for MSVM. This implies that the offline method can exploit the information conveyed by the SFs more efficiently. The same is also evident from SFs accuracy gain in Fig.~\ref{FIG:Model_vs_global}, showing the difference between the experimental MIA without SFs and with only SFs, while MSVM is benefiting more, especially for high INR. With this respect, we also observe in practice the phenomenon of limited accuracy improvement for $\gamma_T \leq 15 \mskip3mu$dB predicted by the error model in \eqref{EQ:ErFunction}.
\begin{figure}
\centering
\includegraphics[width=0.48\textwidth,clip, trim=0cm 0cm 0cm 0cm]{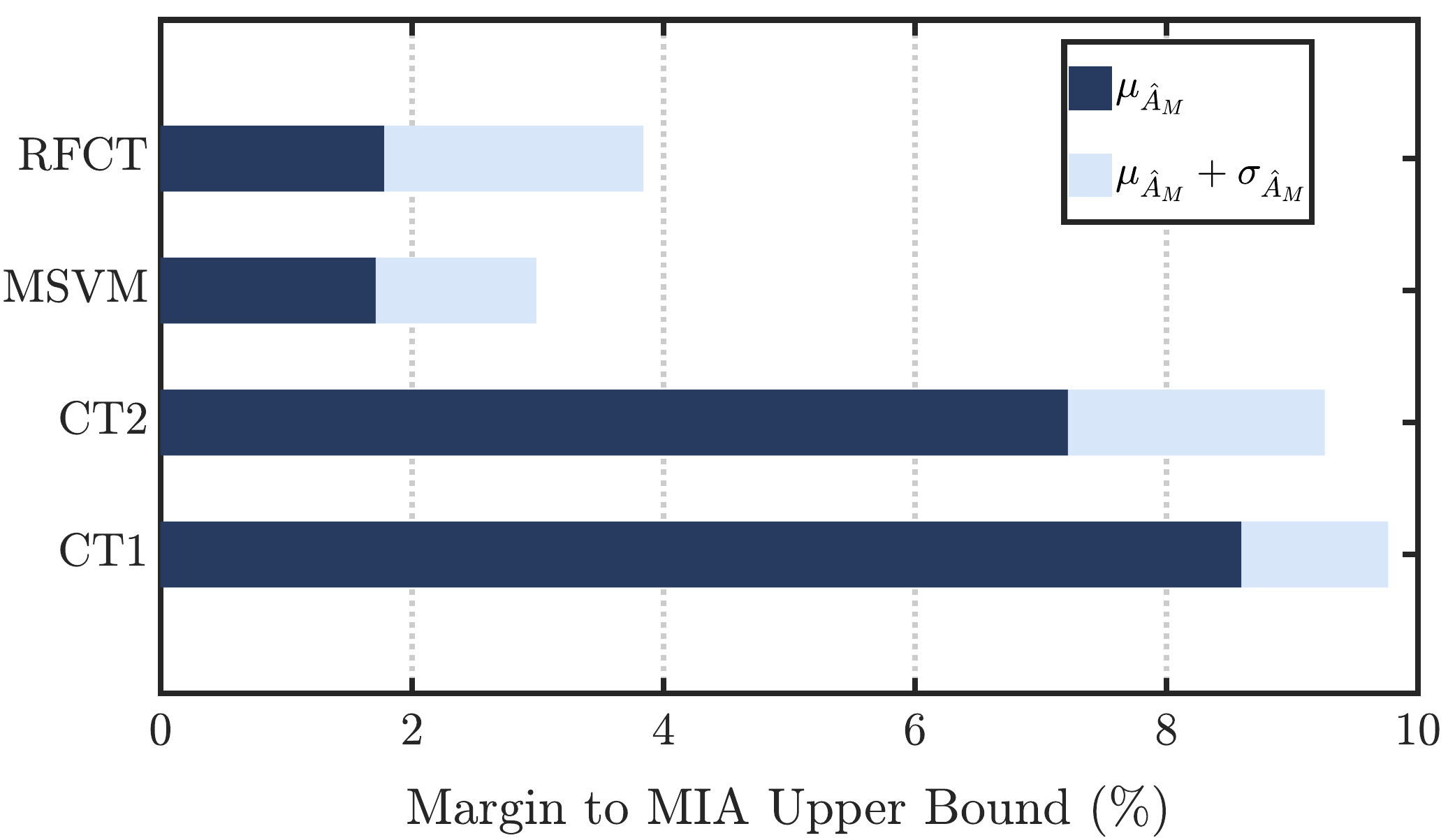}
\caption{Capability of the different classifiers in approaching the estimated upper bound of mean identification accuracy (MIA).}
\label{FIG:classifier_margin_UMI}
\vspace{-12pt}
\end{figure}

To investigate how close the each proposed IDI variant can get to the theoretical MIA upper-bound obtainable with SFs, we use the estimator \eqref{EQ:model_estimator} with all the candidate classification methods. In Fig.~\ref{FIG:classifier_margin_UMI}, we show the difference of the experimental MIA from the upper-bound in terms of mean and standard deviation. It shows that there is a gap of 5$\mskip3mu\%$ between online and offline methods, which highlights different capability to exploit the SFs information due to the structure of the different classifiers. In particular, MSVM and RFCT perform equally well, ensuring extremely limited $<2 \mskip3mu\%$ gap from the theoretical bound. It suggests that, with complex classification strategies approaching to maximum performance, the improvement in IDI performance must be addressed by augmented feature space or an enhanced observation system.

\section{A Use Case}
\label{SEC:Discussion1}
\subsection{Real-Time Estimation of Interference Traffic Distribution}
The inherent capability of the proposed IDI in isolating bursts from multiple concurrent IRNs opens up new opportunities for coexistence modeling and enhancement~\cite{Aamir_PCM, Petrova_IdleTime}. Although devising a coexisting strategy is not in the scope of this work, we demonstrate IDI’s effectiveness in extracting the traffic statistics of an interference traffic that is interweaved with other concurrent heterogeneous interference. To this end, by employing the experimental setup E4, we expose the WSN node to 802.11 and 802.15.1 interference. The node senses a fixed channel $c_Z$ and by using IDI autonomously isolates interference-specific traffic i.e., packet interarrival-time cumulative distribution function (IT-CDF) or packet on-air-time CDF (OAT-CDF), alternatively.

\subsubsection{Real-Time Estimation of IT-CDF}
\begin{figure}
\centering
\includegraphics[width=0.48\textwidth,clip, trim=0cm 0cm 0cm 0cm]{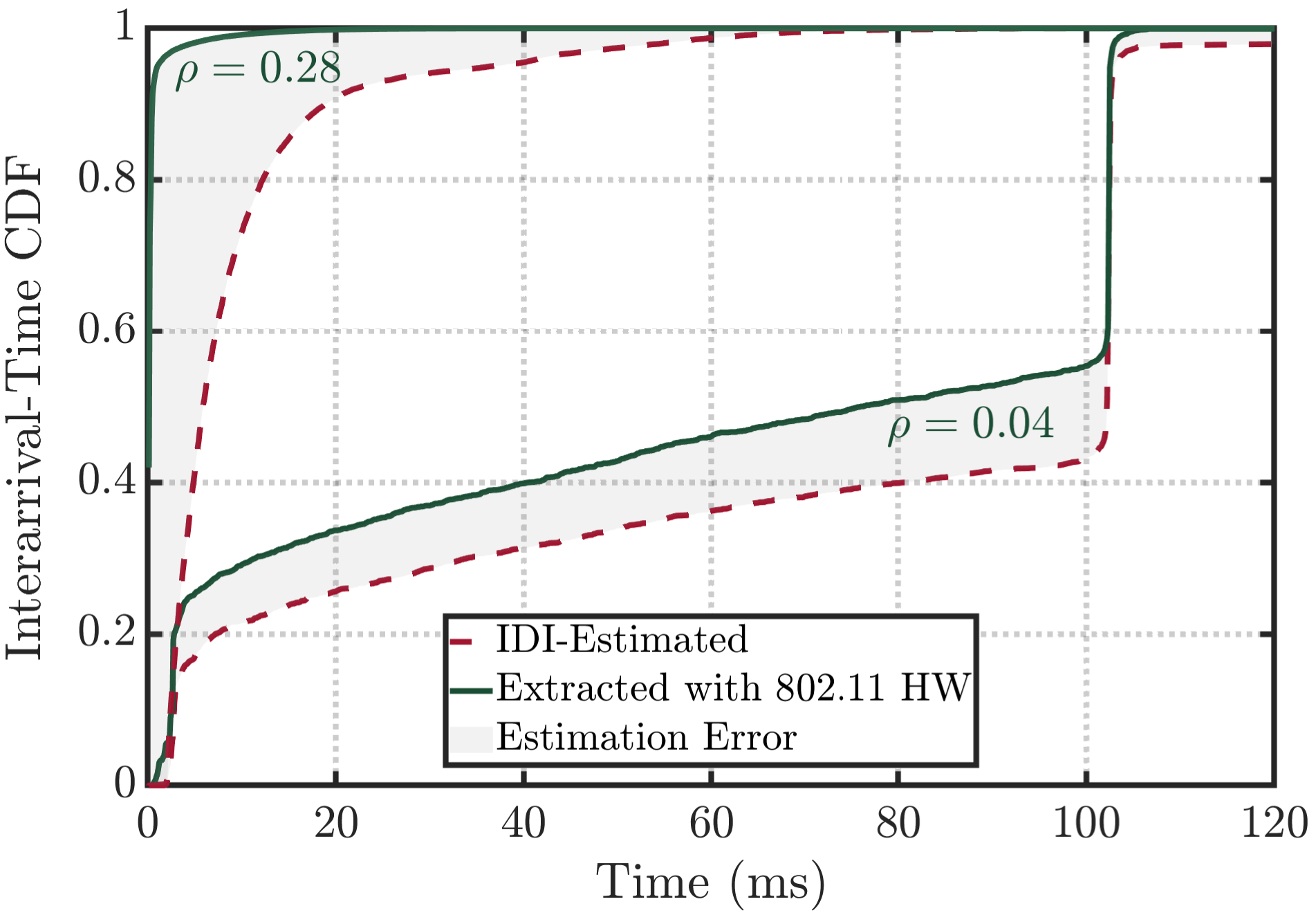}
\caption{Estimated vs 802.11 HW-extracted IT-CDF for 802.11g network with different activity factors $\rho$ under mutual 802.15.4, 802.15.1, and BLE interference.}
\label{FIG:interarrival_wifi}
\vspace{-12pt}
\end{figure}

\begin{figure}
\centering
\includegraphics[width=0.5\textwidth,clip, trim=0cm 0cm 0cm 0cm]{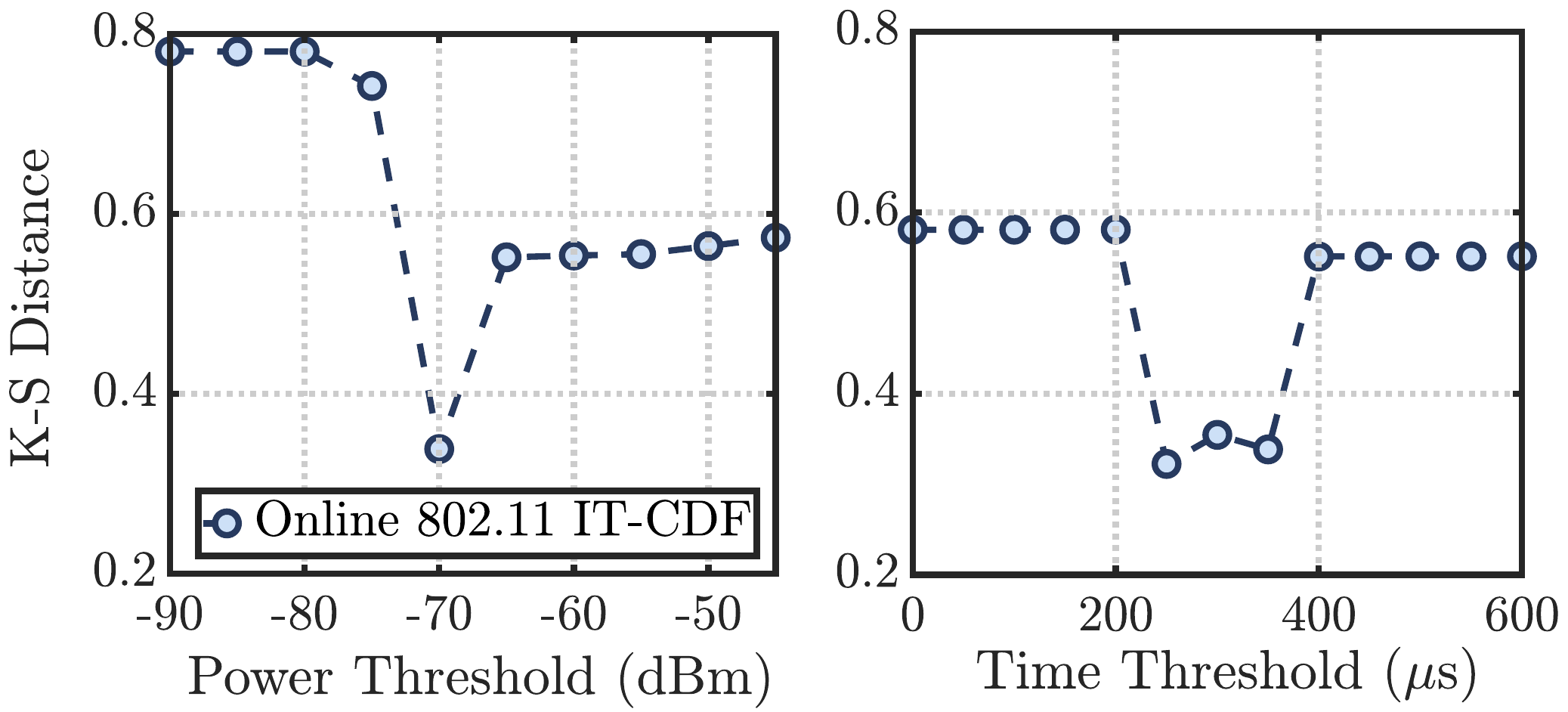}
\caption{Kolmogorov-Smirnof distance between estimated and measured IT-CDF of 802.11 packets, as function of minimum power and packet duration thresholds.}
\label{FIG:kolmogorov}
\vspace{-12pt}
\end{figure}

Fig.~\ref{FIG:interarrival_wifi} compares the estimated IT-CDF---using IDI---and accurately measured IT-CDF---using Wireshark---of a 802.11 IRN for different channel activity factors $\rho$.  It shows that, despite the contrasting capability of the two measurement systems, IDI is able to estimate a reasonably close representation of IT-CDF. To quantify the closeness of the measured and estimated IT-CDFs, we adopt Kolmogorov-Smirnov (K-S) distance measure $D_K$, defined as $D_K(F_A(x),F_B(x))= \sup_x |F_A(x)-F_B(x)|$. We utilize the K-S to find the parameters (the RSSI level and packet OAT) that minimize the $D_K$ distance between the measured and estimated CDFs. It also serves as an additional benchmark for the detection performance of our IDI. The results, as shown in Fig.~\ref{FIG:kolmogorov}, indicate that $D_K$ is minimum when the minimum RSSI and OAT are -70$\mskip3mu$dBm and 250$\mskip3mu\mu$s--350$\mskip3mu\mu$s, respectively. These results clearly reflect the physical limits of the COTS-platforms: i.e., 1) the different sensitivity and frequency response of the 802.15.4 radio, showing $\approx 20 \mskip3mu$dB offset with respect to the 802.11 interface and, 2) the constraint on 324$\mskip3mu\mu$s on minimum OAT of identifiable bursts, as shown in Table~\ref{TAB: accuracy_global}, due to the minimum time required to extract at least one SF.

\begin{figure}
    \centering
  \subfloat[CSMA-CA MAC.]{%
       \includegraphics[width=0.48\linewidth,clip, trim=0cm 0cm 0cm 0cm]{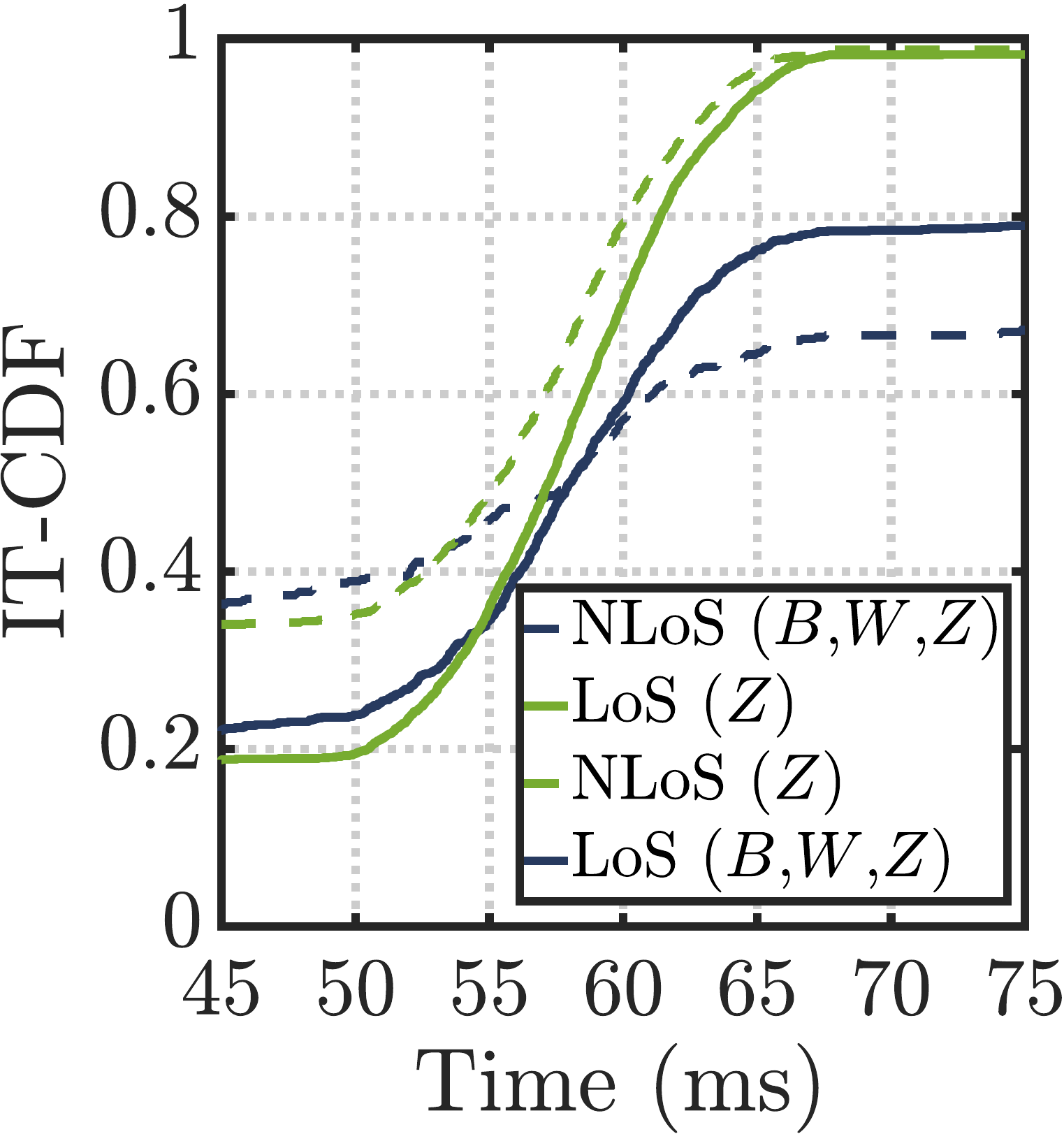}\label{fig:IT_Zigbee_CSMA}}\hfill
  \subfloat[TDMA MAC.]{%
        \includegraphics[width=0.48\linewidth,clip, trim=0cm 0cm 0cm 0cm]{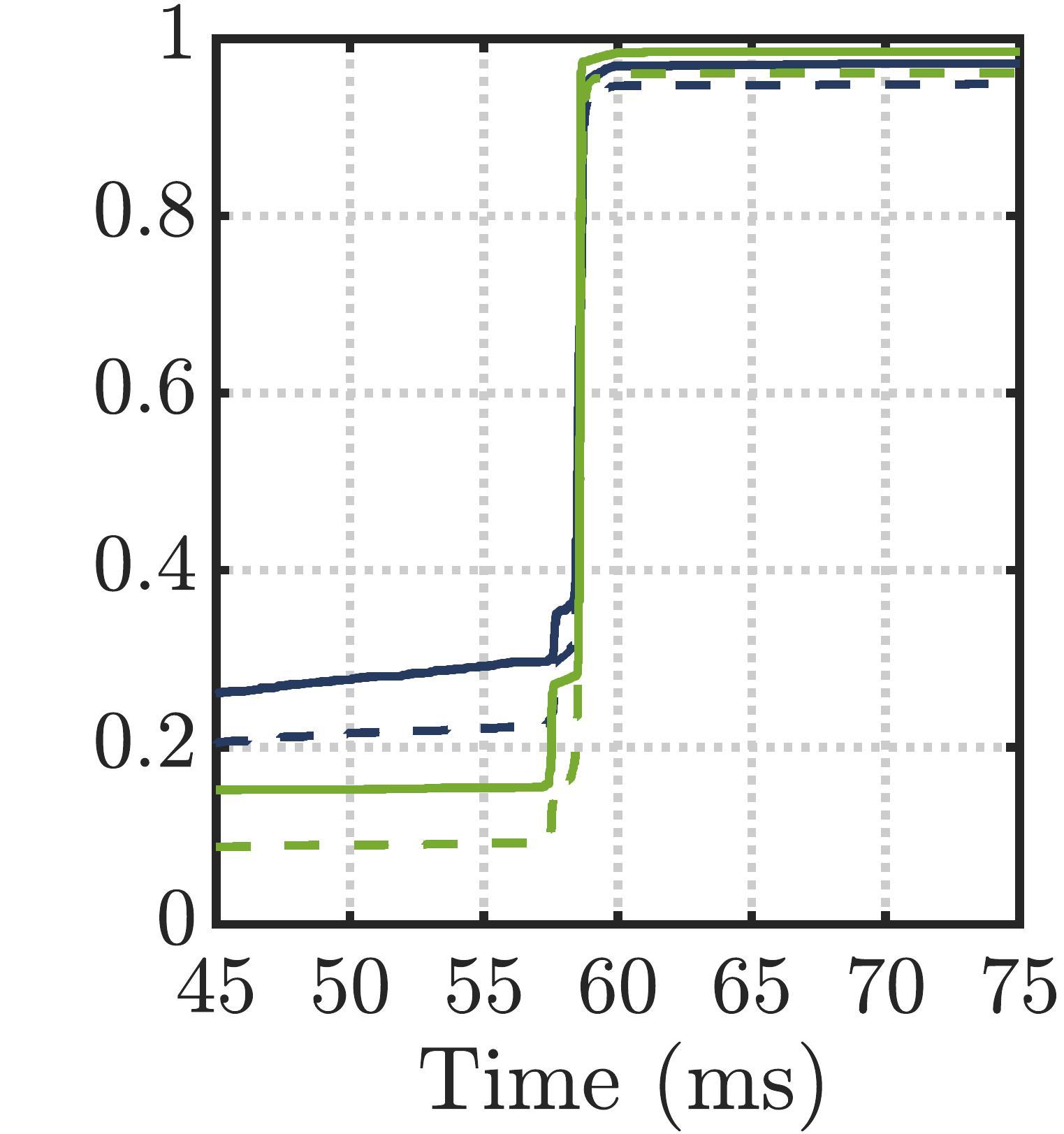}\label{fig:IT_Zigbee_TDMA}}              
  \caption{Estimated IT-CDF for 60$\mskip3mu$ms-periodic 802.15.4 packets transmitted with different MAC protocols under heavy interference ($B,W,Z$) and no concurrent interference ($Z$).}
  \label{fig:IT_Zigbee}
	\vspace{-12pt}
\end{figure}

An IT-CDF is also estimated for the 802.15.4 traffic, where a WSN node periodically generates 32$\mskip3mu$B-payload packets every 60$\mskip3mu$ms on a specific $c_Z$ channel. The results of two experiments showing the estimated IT-CDF for nodes operating with different retransmission strategies are available in Fig.~\ref{fig:IT_Zigbee}, including different LoS condition and two interference scenarios. The employed MAC methods are a CCA-enabled strategy with random back-off retransmission, and a TDMA approach, meaning that the transmission happens every 60$\mskip3mu$ms but in pre-determined slots only. The effects of the heavily crowded spectrum are visible in the IT-CDF in Fig.~\ref{fig:IT_Zigbee_CSMA} suggesting frequent CCA-fails, while the result from interference-free scenario shows a steeper CDF, indicating that that the effect of retransmission is less severe.   

\subsubsection{Real-Time Estimation of OAT-CDF}
\begin{figure}
\centering
\includegraphics[width=1\linewidth]{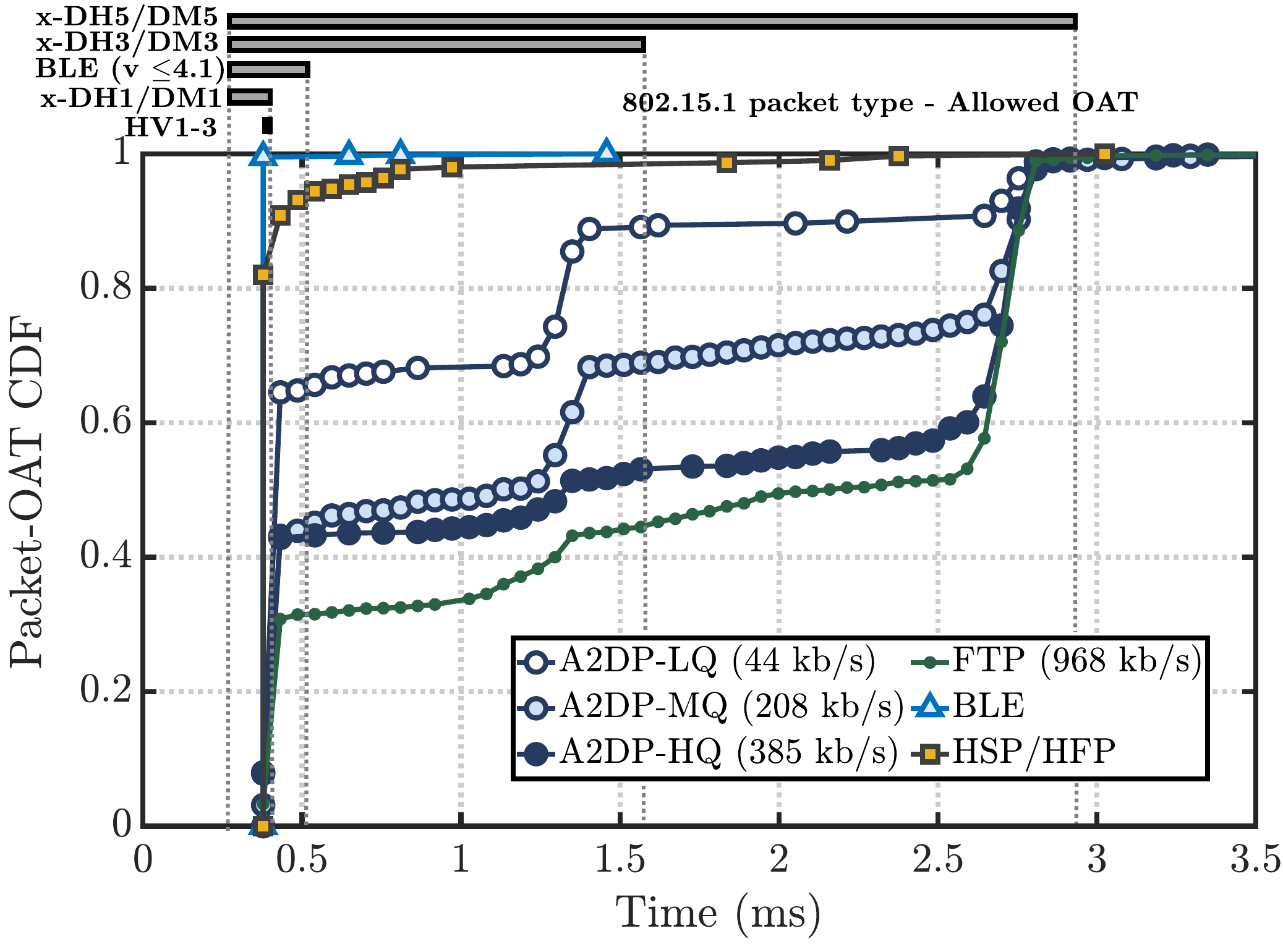}
\caption{Estimated OAT-CDF for 802.15.1 and BLE for different service profiles, under mutual 802.11 and 802.15.4 interference.}
\label{FIG:Bluetooth_CDF}
\vspace{-12pt}
\end{figure}

The estimation of IT-CDF is simplified when IRN operates on a fixed channel. However due to FFH employed in 802.15.1, where a new channel $c_B$ is selected every 625$\mskip3mu\mu$s in a pseudo-random manner, the estimation of interference IT-CDF is non-trivial. Therefore, we leave the study of an intelligent scanning strategy for frequency hopping systems as a future work. Whereas, in this work, we employ the IDI with a simple linear-spectrum channel scanning strategy and collect statistic on the 802.15.1 packet lengths. Fig.~\ref{FIG:Bluetooth_CDF}  shows the estimated CDF for  a number of 802.15.1 applications, including high quality audio-streaming (employing A2DP profile with different bit-rates), headset-quality audio streaming (HSP profile), file transfer, and BLE beacons.

We compare the estimated OAT-CDF in the light of \textit{a priory} known range of OAT of 802.15.1 packets. For instance, file transfer and audio streaming mainly employ high data-rate packets DH1, DH3 and DH5 which have maximum OAT of $[1,3,5] \times$625$\mskip3mu\mu$s respectively, while HSP and BLE beacons are on the threshold of identifiability, due to short OAT packets. In general, we observe that the applications requesting high-bit rates usually utilize the multi-slot packets and thus result in higher channel occupancy. Although the extraction of IT-CDF for the 802.15.1 interference appears rather unrealistic, these results show that the proposed IDI is able to reasonably infer the nature of traffic generated by the IRN in the heavily interfered environment, suggesting employment in mechanisms for coexistence and interference mitigation.

\section{Conclusion}
\label{SEC:Conclusion}
In this paper, we developed a proactive yet lightweight interference detection and identification scheme for COTS radios, which can empower a network of battery-operated devices to coexist reliably with other interfering networks in massive IoT environments. In the design process, we secure that the routine network operation remains unaffected by exploring the trade-offs between performance and complexity. To this end, first, we propose the powerful but lightweight signal features by exploiting the physical properties of the target interference signals and the employed observation system. Second, we carefully investigate a scalable supervised-learning (SL) identification system such that the level of complexity remains under strict control. We force this by targeting an implementable solution for a 15-years old WSN-hardware. In performance, the proposed method enables autonomous and real-time detection and identification of signal bursts, showing $>90 \mskip3mu\%$ accuracy even in heavily multi-technology interference scenarios. We employ offline machine-learning methods as a reference benchmark, showing that our online implementation exhibit very limited performance gap, while ensuring 20-fold shorter processing delay. What makes our solution gaining in identification accuracy are the innovative spectral features, which we backed up by developing an analytical model. This model is utilized to estimate an upper bound on the identification gains, and to show the efficiency of the proposed method in exploiting the available spectral information. Finally, a realistic use-case is shown by means of an autonomous system capable of isolating and estimating traffic distributions of concurrent heterogeneous interfering networks. Our system is a strong candidate for real-time adoption in many existing cognitive approaches hunting channel idle-times for transmissions.

\bibliographystyle{IEEEtran}
\bibliography{Bibliography}

\end{document}